\documentclass[11pt]{article}
\usepackage[left=1in,top=1in,right=1in,bottom=1in,head=.1in,nofoot]{geometry}

\usepackage{setspace,url,bm,amsmath} 

\usepackage{titlesec} 
\titlelabel{\thetitle.\quad} 

\usepackage[margin=20pt]{subcaption}

\usepackage{amsmath,amsfonts,amsthm,amssymb}
\usepackage{graphicx}

\usepackage[colorlinks=true,linkcolor=black,citecolor=blue,urlcolor=blue]{hyperref}
\usepackage{threeparttable}
\usepackage{manyfoot}%
\usepackage{booktabs}%
\usepackage{algorithm}%
\usepackage{algorithmicx}%
\usepackage{algpseudocode}%
\usepackage{listings}%
\usepackage{dsfont}%
\usepackage{bm}
\usepackage{etoolbox}

\usepackage{comment}
\theoremstyle{definition}
\newtheorem{Assumption}{Assumption}
\newtheorem*{theorem*}{Theorem}
\newtheorem{theorem}{Theorem}
\newtheorem{proposition}{Proposition}
\newtheorem{lemma}{Lemma}
\newtheorem{remark}{Remark}

\newtheorem*{corollary*}{Corollary}

\DeclareFontFamily{U}{mathx}{\hyphenchar\font45}
\DeclareFontShape{U}{mathx}{m}{n}{
      <5> <6> <7> <8> <9> <10>
      <10.95> <12> <14.4> <17.28> <20.74> <24.88>
      mathx10
      }{}
\DeclareSymbolFont{mathx}{U}{mathx}{m}{n}
\DeclareFontSubstitution{U}{mathx}{m}{n}
\DeclareMathAccent{\widecheck}{0}{mathx}{"71}
\DeclareMathAccent{\wideparen}{0}{mathx}{"75}

\usepackage{float}
\usepackage{natbib}
\usepackage{indentfirst}

\allowdisplaybreaks[4]

\def\T{\text{T}}
\def\Var{\text{Var}}

\def\tauor{\hat \tau_{\text{oracle}}}
\def\taupro{\hat \tau_{\text{emp}}}
\def\tauss{\hat \tau_{\text{ss}}}

\def\taulin{\hat \tau_{\text{linear}}}
\def\tauker{\hat \tau_{\text{kernel}}}
\def\tauns{\hat \tau_{\text{nspline}}}
\def\taurf{\hat \tau_{\text{rf}}}
\def\taunn{\hat \tau_{\text{nnet}}}

\def\ttaulin{\tilde \tau_{\text{linear}}}
\def\ttauker{\tilde \tau_{\text{kernel}}}
\def\ttauns{\tilde \tau_{\text{nspline}}}
\def\ttaurf{\tilde \tau_{\text{rf}}}
\def\ttaunn{\tilde \tau_{\text{nnet}}}

\def\taulasso{\hat \tau_{\text{lasso}}}
\def\taunnet{\hat \tau_{\text{nnet}}}
\def\taugbm{\hat \tau_{\text{gbrt}}}
\def\taurfh{\hat \tau_{\text{rf}}}
\def\taurpart{\hat \tau_{\text{rpart}}}

\def\ttaulasso{\tilde \tau_{\text{lasso}}}
\def\ttaunnet{\tilde  \tau_{\text{nnet}}}
\def\ttaugbm{\tilde  \tau_{\text{gbrt}}}
\def\ttaurfh{\tilde  \tau_{\text{rf}}}
\def\ttaurpart{\tilde  \tau_{\text{rpart}}}

\def\taulassoss{\hat \tau_{\text{lasso}}^{\text{ss}}}
\def\taunnetss{\hat \tau_{\text{nnet}}^{\text{ss}}}
\def\taugbmss{\hat \tau_{\text{gbrt}}^{\text{ss}}}
\def\taurfhss{\hat \tau_{\text{rf}}^{\text{ss}}}
\def\taurpartss{\hat \tau_{\text{rpart}}^{\text{ss}}}

\def\ttaulassoss{\tilde \tau_{\text{lasso}}^{\text{ss}}}
\def\ttaunnetss{\tilde  \tau_{\text{nnet}}^{\text{ss}}}
\def\ttaugbmss{\tilde  \tau_{\text{gbrt}}^{\text{ss}}}
\def\ttaurfhss{\tilde  \tau_{\text{rf}}^{\text{ss}}}
\def\ttaurpartss{\tilde  \tau_{\text{rpart}}^{\text{ss}}}

\def\sumi{\sum_{i=1}^{n}}
\def\sumj{\sum_{j=1}^{n}}
\def\sumk{\sum_{k=1}^{K} }
\def\sumik{\sum_{i \in [k]}}
\def\sumjk{\sum_{j \in [k]}}

\def\nkt{n_{[k]1}}
\def\nkc{n_{[k]0}}

\def\pink{\pi_{n[k]}}

\def\pnk{p_{n[k]}}

\def\pk{p_{[k]}}

\def\nk{n_{[k]}}

\def\nt{n_1}
\def\nc{n_0}

\def\YkThat{\bar{Y}_{[k]1}}
\def\YkChat{\bar{Y}_{[k]0}}

\def\bx{X}

\def\XkT{\bar{\bx}_{[k]}}
\def\XkC{\bar{{\bx}}_{[k]}}
\def\XkThat{\bar{{\bx}}_{[k]1}}
\def\XkChat{\bar{{\bx}}_{[k]0}}

\def\pmi{(1-\pi)}
\def\hk{h_{[k]}(X)}

\def\hax{h_{[k]}(\cdot, a)}

\def\hathts{\hat h_{[k]}(\bx_i, 1)}
\def\hathcs{\hat h_{[k]}(\bx_i, 0)}
\def\hathtms{\hat h_{[k]m}(\bx_i, 1)}
\def\hathcms{\hat h_{[k]m}(\bx_i, 0)}
\def\hatha{\hat h_{[k]}(\bx_i, a)}

\def\hts{h_{[k]}(\bx_i, 1)}
\def\hcs{h_{[k]}(\bx_i, 0)}
\def\ha{h_{[k]}(\bx_i, a)}
\def\hk{h_{[k]}(\bx_i)}
\def\barhkt{\bar{h}_{[k]1}}
\def\barhkc{\bar{h}_{[k]0}}

\def\barhathat{ \bar{\hat h}_{[k]1}(\cdot, a)}
\def\barhat{\bar{h}_{[k]1}(\cdot, a)}
\def\barhathac{ \bar{\hat h}_{[k]0}(\cdot, a)}
\def\barhac{\bar{h}_{[k]0}(\cdot, a)}

\def\mhxq{\hat{m}_H(x)}
\def\mxq{m(x)}

\def\pinkm{\pi_{n[k]m}}
\def\pinkma{\pi_{n[k]ma}}
\def\Tnkm{T_{n[k]m}}
\def\bTnkm{\breve{T}_{n[k]m}}
\def\Wk{W_{[k]}}

\def\Nkam{\mathbb{N}_{[k]ma}}
\def\nkam{n_{[k]ma}}
\def\nkm{n_{[k]m *}}
\def\nktm{n_{[k]m1}}
\def\nkcm{n_{[k]m0}}
\def\Dika{\Delta_{i[k]ma}}
\def\bDika{\breve{\Delta}_{i[k]ma}}
\def\sumnkam{\sum_{i=1}^{\nkam}}
\def\sumnkamc{\sum_{i=\nkam + 1}^{\nkm}}
\def\sumkm{\sum_{i\in [k] \cap I_m}}
\def\sumjkm{\sum_{j\in [k] \cap I_m}}
\def\summ{\sum_{m=1}^M}
\def\pnkm{p_{n[k]m}}
\def\nm{n_m}
\def\hatham{\hat h_{[k]m}(\bx_i, a)}
\def\barhathatm{ \bar{\hat h}_{[k]m1}(\cdot, a)}
\def\barhathacm{ \bar{\hat h}_{[k]m0}(\cdot, a)}
\def\barhatm{\bar{h}_{[k]m1}(\cdot, a)}
\def\barhacm{\bar{h}_{[k]m0}(\cdot, a)}

\def\brktm{\bar{\hat{r}}_{[k]m1}}
\def\brkcm{\bar{\hat{r}}_{[k]m0}}
\def\brtm{\bar{\hat{r}}_{m1}}
\def\brcm{\bar{\hat{r}}_{m0}}
\def\pin{\pi_{n}}
\def\pim{\pi_{nm}}

\title{\bf  A unified framework for covariate adjustment under stratified randomization}
\date{}

\author{
\small
{
Fuyi Tu$^{1,2}$, \ \ Wei Ma$^{2}$, \ \ Hanzhong Liu$^{3}$\thanks{\small{Correspondence: \texttt{lhz2016@tsinghua.edu.cn}}}
}
\\ \\
{\small $^{1}$ School of Science, Chongqing University of Posts and Telecommunications, Chongqing, China}\\
{\small $^{2}$ Institute of Statistics and Big Data, Renmin University of China, Beijing, China}\\
{\small $^{3}$ Center for Statistical Science, Department of Industrial Engineering, Tsinghua University, Beijing, China}
}

\begin{document}
\doublespacing

\maketitle


\begin{abstract}

Randomization, as a key technique in clinical trials, can eliminate sources of bias and produce comparable treatment groups. In randomized experiments, the treatment effect is a parameter of general interest. Researchers have explored the validity of using linear models to estimate the treatment effect and perform covariate adjustment and thus improve the estimation efficiency. However, the relationship between covariates and outcomes is not necessarily linear, and is often intricate. Advances in statistical theory and related computer technology allow us to use nonparametric and machine learning methods to better estimate the relationship between covariates and outcomes and thus obtain further efficiency gains. However, theoretical studies on how to draw valid inferences when using nonparametric and machine learning methods under stratified randomization are yet to be conducted. In this paper, we discuss a unified framework for covariate adjustment and corresponding statistical inference under stratified randomization and present a detailed proof of the validity of using local linear kernel-weighted least squares regression for covariate adjustment in treatment effect estimators as a special case. In the case of high-dimensional data, we additionally propose an algorithm for statistical inference using machine learning methods under stratified randomization, which makes use of sample splitting to alleviate the requirements on the asymptotic properties of machine learning methods. Finally, we compare the performances of treatment effect estimators using different machine learning methods by considering various data generation scenarios, to guide practical research.

\vspace{12pt}
\end{abstract}

\section{Introduction}

As a method of preventing selection bias and producing valid statistical inference, randomization has been extensively used in clinical trials. Although simple randomization tends to balance both known and unknown covariates on average, as stated in \cite{Cornfield1959}, severe imbalances of important covariates among treatment groups may still occur in practice. Researchers seek methods that better balance a few pre-specified important covariates. Various stratified randomization methods, such as stratified block randomization \citep{Zelen1974}, stratified biased coin randomization \citep{Efron1971}, and minimization \citep{Taves1974, Pocock1975}, have been proposed. In stratified randomization, a set of covariates is used to form strata, and subjects are then assigned to treatment groups with a probability related to their strata. Stratified randomization further improves the imbalance of important covariates among groups and yields more acceptable and efficient estimators, especially under small sample sizes.

It has been widely suggested in the statistical literature that covariates used in the randomization stage also be included in the analysis stage, as otherwise tests may be invalid \citep{Kahan2012}. Stratified analysis provides an aggregate test over strata by pooling the analysis results of all the strata. It takes the interaction between outcomes and strata induced by stratified randomization into account, and leads to valid and efficient inference. As in most randomized controlled trials, we focus on the inference of the average treatment effect, which characterizes the effect of a typical treatment on subjects. \cite{Bugni2018,Bugni2019} studied the asymptotic property of the naive difference-in-means estimator and the ordinary-least-squares (OLS) estimator obtained from regressing outcomes on indicators of strata under stratified randomization. However, as proposed by \cite{Ma2020Regression,Liu2020Lasso} and \cite{Ye2020Inference}, adjusting for additional covariates through linear regressions, such as OLS or lasso \citep{Tibshirani1996}, further improves the efficiency of the treatment effect estimator. The variance is reduced by projecting outcomes onto the space spanned by linear functions of covariates and further analyzing the residuals. When there is no strong evidence of a linear relationship between covariates and outcomes, the aforementioned projection may be limited in improving the efficiency of the treatment effect estimator. \cite{Wang2021} adapted the adjusting methods to parametric regressions, treated the estimator of the parameter of interest as an M-estimator and studied its properties.

As pointed out by \cite{Tsiatis2008}, we can consider using nonparametric and machine learning methods to further improve the efficiency of treatment effect estimators, especially in the presence of high-dimensional covariates. This direction of research has been explored in recent years. For example, \cite{Williams2021} evaluated the performance of four machine learning methods for ordinal and time-to-event outcomes and demonstrated that the use of machine learning methods generally improves the estimation efficiency. In particular, the estimation efficiency can be greatly increased by having a sufficient number of samples. Additionally, some evidence shown in \cite{Liu2020Lasso} showed that consistent and efficient estimates can be obtained when reasonably using covariate information in high-dimensional cases.

However, drawing valid inferences when using nonparametric methods and machine learning methods under stratified randomization is challenging, especially in the case of high-dimensional data.
In this paper, we first derive the asymptotic property of an ``oracle'' estimator, where the true projection function of the outcome on the space spanned by arbitrary functions of covariates is plugged in and the estimator is thus expected to perform the best. We then outline the restriction on the convergence rate of the estimation of the true projection function, under which we can obtain an valid empirical treatment effect estimator. Additionally, we discuss the optimal choice of the projection function in the sense of treatment effect estimator's efficiency, and provide insights into the use of different covariate adjustment tools, based on a certain degree of theoretical justification.
Next, we suggest the use of statistical methods to obtain a consistent estimation of the optimal projection function. When the covariates have low dimensionality, we can use general nonparametric methods such as the local linear kernel-weighted least squares regression (abbreviated as local linear kernel in the following text); when the covariates have relatively high dimensionality, we can use the lasso or machine learning methods, such as random forest \citep{Breiman2001}, neural networks, and debiased/double machine learning methods \citep{Chernozhukov2018}. Finally, we report on the evaluation of the empirical performances of the proposed estimators using different estimating methods for the projection function are evaluated in a simulation study.

\section{Framework and notation}
We consider an experiment with two treatments under stratified randomization and follow the framework and notation of \cite{Ma2020Regression} and \cite{Liu2020Lasso}. Let $A_i, i = 1,\dots,n$, denote the indicator for treatment assignment, i.e., $A_i = 1$ if the $i$th unit is assigned to the treatment group and $A_i = 0$ otherwise. The target proportion of treated units is $\pi = P(A_i = 1) \in (0,1)$. We assume that the observed outcome $Y_i$ is a function of treatment assignment $A_i$ and potential outcomes under treatment ($Y_i(1)$) and control ($Y_i(0)$): $Y_i = A_i Y_i(1) + (1-A_i) Y_i(0)$. During the experiment, units are stratified into $K$ strata, with $B_i \in \{1, \dots, K\}$ denoting the specific stratum that unit $i$ falls into. To rule out the empty stratum, we assume that the probability of units being assigned to each stratum is positive, i.e., $\pk = P(B_i = k) > 0$, for $k \in \{1,\dots, K\}$ and $i \in \{1,\dots,n\}$. Additionally, we collect a $p$-dimensional vector of baseline covariates for each unit, denoted as $X_i = (X_{i1},\dots,X_{ip})^\textnormal{T}$. The covariates can be either low-dimensional or high-dimensional, and the estimation method is changed accordingly. We use the subscript $1$ or $0$ to indicate the assigned group being the treatment group or control group. In the randomized experiment, $n_1 = \sum_{i=1}^n A_i$ units are assigned to the treatment group and $n_0 = \sum_{i=1}^n (1-A_i)$ units are assigned to the control group. Moreover, we use the subscript $[k]$ to index the statistics in stratum $k$, e.g., the number of units, treated units, and control units in stratum $k$ are $\nk = \sumik 1, \nkt = \sumik A_i$, and $\nkc = \sumik (1-A_i)$, respectively, where $i \in [k]$ indexes units in stratum $k$. The proportion of stratum sizes and the treated units in stratum $k$ are denoted as $\pnk = \nk/n$ and $\pink = \nkt/\nk$, respectively. Our parameter of interest is the average treatment effect: $\tau = E\{Y_i(1) - Y_i(0)\}$.

Let $\mathcal{R}_2 = \{V:\max_{k=1,\dots,K} \textnormal{Var}\{V|B_i = k\}>0\}$ be the set of random variables with at least one positive stratum-specific variance. Throughout this paper, the following requirements are imposed on the data-generating process and randomization mechanism.

\begin{Assumption}\label{assum1}
  $\{Y_i(1), Y_i(0), X_i\}_{i=1}^n$ are independent and identically distributed (i.i.d.) samples from the population distribution of $\{Y(1), Y(0), X\}$. Moreover, $E\{Y_i^2(a)\} < \infty, Y_i(a) \in \mathcal{R}_2, a = 0, 1$.
\end{Assumption}
\begin{Assumption}\label{assum2}
  Conditional on $B^{(n)} = \{B_1,\dots,B_n\}$, treatment assignments $A^{(n)} = \{A_1,\dots,A_n\}$ and $\{Y_i(1),Y_i(0), X_i\}_{i=1}^n$ are independent.
\end{Assumption}
\begin{Assumption}\label{assum3}
  In each stratum, the proportion of treated units $\pink$ converges in probability to $\pi$.
\end{Assumption}

The above assumptions were also made by \cite{Bugni2019}, \cite{Ma2020Regression} and \cite{Liu2020Lasso}. Here, we state no restrictions on the relationship between $A_i$, indicating that they can be related to each other, which is a common case under stratified randomization. When applying different estimation methods to baseline covariates, additional assumptions are made for $X_i$. Assumption~\ref{assum2} holds under simple and restricted randomization \citep{Rosenberger2015}, and is widely satisfied by existing stratified randomization methods and covariate-adaptive randomization methods, including the stratified biased-coin design \citep{Efron1971}, stratified block design \citep{Zelen1974}, stratified adaptive biased-coin design \citep{Wei1978}, Pocock and Simon's minimization \citep{Pocock1975}, and the methods proposed by \cite{Hu2012}.

{\bf Notation}. The mean and variance of a random variable $V$ are denoted as $\mu_V = E(V)$ and $\sigma_V^2 = \textnormal{Var}(V)$, respectively. Let $\tilde{V} = V - E(V|B)$ be the variable $V$ centered at its stratum-specific mean. For random variables $r_i(a), i = 1,\dots,n, a \in \{0,1\}$, the sample means under treatment and control are denoted as $\bar{r}_1 = (1/n_1)\sumi A_ir_i(1)$ and $\bar{r}_0 = (1/n_0)\sumi(1-A_i)r_i(0)$, respectively, and the stratum-specific sample means are denoted as $\bar{r}_{[k]1} = (1/\nkt)\sumik A_ir_i(1)$ and $\bar{r}_{[k]0} = (1/\nkc)\sumik (1-A_i)r_i(0)$, respectively. The variance of the treatment effect estimator consists of the variation of (transformed) potential outcomes $r_i(1)$ and $r_i(0)$ and the sum of variations from treatment effect heterogeneity within each stratum:
{\small$$\varsigma_r^2(\pi) = \frac{1}{\pi}\sigma^2_{r_i(1)-E\{r_i(1)|B_i\}} + \frac{1}{1-\pi}\sigma^2_{r_i(0)-E\{r_i(0)|B_i\}},$$
$$\varsigma_{Hr}^2 = \sumk  \pk \Big(\big[E\{r_i(1)|B_i=k\} - E\{r_i(1)\}\big] - \big[E\{r_i(0)|B_i=k\} - E\{r_i(0)\}\big]\Big)^2.$$}
We express the sample analog as
{\small\begin{eqnarray}
\hat \varsigma^2_{r}(\pi) &= & \frac{1}{\pi}  \sumk \pnk  \bigg[  \frac{1}{\nkt} \sumik  A_i  \Big\{  \hat r_i(1)  - \frac{1}{\nkt} \sumjk A_j  \hat r_j(1)  \Big\} ^2  \bigg] \nonumber\\
&& + \frac{1}{1 - \pi} \sumk \pnk  \bigg[  \frac{1}{\nkc} \sumik (1 - A_i ) \Big\{ \hat r_i (0)  -   \frac{1}{\nkc} \sumjk ( 1 - A_j )  \hat r_j (0)  \Big\}^2   \bigg], \nonumber
\end{eqnarray}}
{\small
\begin{eqnarray*}
\hat  \varsigma^2_{H r} &= &\sumk \pnk  \bigg[ \Big\{ \frac{1}{\nkt} \sumjk A_j  \hat r_j(1)   - \frac{1}{\nt} \sumi A_i \hat r_i (1) \Big\}  \\
& &  -  \Big\{  \frac{1}{\nkc} \sumjk ( 1 - A_j )  \hat r_j (0) - \frac{1}{\nc}  \sumi ( 1 - A_i ) \hat r_i(0)   \Big\} \bigg]^2,
\end{eqnarray*}}
where $\hat{r}_i(a)$ is the observed or estimated value of $r_i(a)$. We define the $L_2$ norm of a function $f$ as $||f||_{L_2} = \{E(f^2)\}^{1/2}$.

\section{Linear adjustment}

Under stratified randomization, the average treatment effect can be consistently estimated by aggregating the treatment effect estimates in each stratum, and a naive estimator is
$$\hat{\tau} = \sumk \pnk (\YkThat  - \YkChat).$$

To further improve estimation efficiency, we consider different methods for covariate adjustment. Linear regression, as a concise and highly interpretable model, is widely accepted in practice and its theoretical properties have been intensively studied. Therefore, a number of recent works on statistical inference under randomization have considered linear regressions for covariate adjustment \citep[e.g.,][]{lin2013,Bugni2018, Bugni2019, Liu2020,LiDing2020,Ma2020Regression,Ye2020}.

Let $\bar{X}_{[k]} = n_{[k]}^{-1}\sumik X_i$, \cite{Liu2020Lasso} proposed a general regression-adjusted treatment effect estimator:
{\small
$$\hat{\tau}_{\mathrm{gen}}=\sumk \pnk\left[\left\{\bar{Y}_{[k] 1}-\left(\bar{X}_{[k] 1}-\bar{X}_{[k]}\right)^{\mathrm{T}} \hat{\beta}_{[k]}(1)\right\}-\left\{\bar{Y}_{[k] 0}-\left(\bar{X}_{[k] 0}-\bar{X}_{[k]}\right)^{\mathrm{T}} \hat{\beta}_{[k]}(0)\right\}\right],$$}
where $\hat{\beta}_{[k]}(1)$ and $\hat{\beta}_{[k]}(0)$ are the estimated regression-adjusted vectors for the treatment and control groups, respectively.
This general estimator is applicable to both low-dimensional cases, using OLS, and high-dimensional cases, using lasso, and the asymptotic properties of the corresponding estimators have been derived. In high-dimensional cases, the general estimator can also be extended to the use of Ridge regression \citep{Hoerl1970} or elastic net regression\citep{Zou2005}.

\section{Nonlinear adjustment}
The idea behind the regression-adjusted estimator is to project outcomes onto the space spanned by linear functions of the covariates and then further analyze the residuals to reduce the variance. Treatment effect estimators that are more efficient can be obtained by appropriately estimating the projection of the outcomes on the (potentially nonlinear) space spanned by the covariates. We can therefore use other methods, such as those based on partial linear models, generalized linear models, nonparametric models and machine-learning methods, for covariate adjustment.

\subsection{Oracle estimator}\label{oracle}
We start by considering the oracle case where the space $\Omega$ spanned by functions of $X$ with finite variance is given. Conditional on $B_i = k$, let $h_{[k]}(X,1)$ and $h_{[k]}(X,0)$ denote the projections of $Y(1)$ and $Y(0)$ onto $\Omega$, respectively. Following the idea of the regression-adjusted estimator, our proposed oracle estimator is:
\begin{eqnarray*}\label{eqn::oracle}
  \tauor &=& \sumk  \pnk \Big[\Big\{\YkThat  - \frac{1}{\nkt} \sumik (A_i - \pink)\hts \Big\} \\
  && - \Big\{\YkChat  + \frac{1}{\nkc} \sumik (A_i - \pink)\hcs \Big\}\Big]. \nonumber
\end{eqnarray*}

\begin{remark}
  If we consider $\Omega$ as a space spanned by linear functions of $X$, then $h_{[k]}(X,a) = X^\T \beta_{[k]}(a)$, and we have
\begin{eqnarray*}
  \frac{1}{\nkt} \sumik (A_i - \pink)\hts &=& \frac{1}{\nkt} \sumik (A_i - \pink) X_i^\T \beta_{[k]}(1)  \\
   &=& \bigg(\frac{1}{\nkt} \sumik A_i X_i^\T \ - \frac{\pink}{\nkt}\sumik X_i^\T \bigg) \beta_{[k]}(1) \\
   &=& \XkThat^\T \beta_{[k]}(1) - \frac{1}{\nk} \sumik X_i^\T \beta_{[k]}(1) \\
   &=& (\XkThat - \XkT)^\T \beta_{[k]}(1).
\end{eqnarray*}
Similarly, we have $\nkc^{-1} \sumik (A_i - \pink) \hcs = (\XkChat - \XkC)^\T \beta_{[k]}(0)$. Thus, our proposed estimator is identical to the oracle regression-adjusted stratum-specific estimator proposed by \cite{Liu2020Lasso}.
\end{remark}

Let $r_i(a) = Y_i (a) - [(1-\pi) \hts + \pi \hcs], i \in [k]$, we establish the following proposition for the oracle estimator. The detailed proof is given in Section 2 of the Supplementary Materials.

\begin{proposition}\label{Prop1}
  Suppose that $r_i(a) \in \mathcal{R}_2, \ E\{ h^2_{[k]}(X_i,a) \} < \infty, \ a=0,1, \ k=1,\dots,K$. Under Assumptions~\ref{assum1}--\ref{assum3},
  $$\sqrt{n}(\tauor - \tau) \stackrel{d}\rightarrow \mathcal{N}(0,\varsigma^2_r(\pi) + \varsigma^2_{Hr}).$$
\end{proposition}

\begin{remark}
     If we assume that $\hts$ and $\hcs$ are the same across strata, then we can replace $\hts$ and $\hcs$ by $h(X_i,1)$ and $h(X_i,0)$ in the estimator and transformed outcomes, respectively, and Proposition~\ref{Prop1} still holds.
\end{remark}

\subsection{Empirical estimator}
Plugging in the estimates of $h_{[k]}(X,a)$, our empirical treatment effect estimator adjusting for baseline covariates is defined as \citep{Liu2020Lasso}:
\begin{eqnarray}\label{eqn::emp}
  \taupro &=& \sumk \pnk \Big[\Big\{\YkThat  - \frac{1}{\nkt} \sumik (A_i - \pink)\hathts\Big\} \nonumber \\
  & & - \Big\{\YkChat  + \frac{1}{\nkc} \sumik (A_i - \pink)\hathcs\Big\}\Big],
\end{eqnarray}
where $\hatha,\ a = 0,1$, are projection functions estimated by different methods.

The discussion presented by \cite{Tsiatis2008} implied that estimates of $\ha,\ a = 0, 1$ are desirable if they can estimate $\ha$ well, but no formal conditions were given on the estimation error. To study the asymptotic properties of $\taupro$ under stratified randomization, the following conditions were outlined by \cite{Liu2020Lasso}.

\begin{Assumption}\label{assum4}
For $k = 1,\dots, K$ and $a=0,1$,
\begin{equation}
\label{eqn::goal1}
  \sqrt{n}  \Big[ \big\{ \barhathat - \barhat \big\}  -  \big\{ \barhathac - \barhac \big\}  \Big] = o_P(1),
\end{equation}
\begin{equation}
\label{eqn::goal2}
  \frac{1}{ \nk } \sumik  \Big\{   \hatha - \ha   \Big\}^2 = o_P(1),
\end{equation}
where $\barhathat$ and $\barhathac$ respectively denote the sample means of $\hatha$ in the treatment group and control group within stratum $k$.
\end{Assumption}

The first equation in Assumption~\ref{assum4}, i.e. Equation~\eqref{eqn::goal1}, implies that the difference between the oracle estimator and the empirical estimator is negligible at the rate of $o_P(n^{-1/2})$, the assumption of the similar form has also been used for the regression-adjusted estimation of quantile treatment effects (Assumption (i) in \cite{Jiang2023}). Equation~\eqref{eqn::goal2} allows control of the estimation error of the projection function and serves as a guarantee of consistent variance estimation. Based on this assumption, we can draw valid inference for the average treatment effect. According to the low or high dimensionality of the covariates, a wide range of methods can be applied to estimate $\ha, \ a = 0, 1$. If we consider the space spanned by arbitrary measurable functions of $X$ with finite variance, then nonparametric methods can be applied. Additionally, if we have prior information on the format of $\ha$, we can adopt parametric regressions. For linear adjustments, these conditions have been verified for lasso and linear regressions. For nonlinear adjustments, \cite{Wang2021} considered the M-estimation of the parameter of interest and proved relevant asymptotic properties. However, theoretical properties of covariate adjusted estimators using more general nonparametric or machine learning methods are not clear and yet to be explored, see discussion in Section~\ref{sec::discussion}. Meanwhile, Assumption~\ref{assum4} is a high-level assumption that, regardless of whether the setting is low-dimensional or not, as long as the adjusting method satisfies this assumption, we can obtain a consistent treatment effect estimator and make valid inference. One of the main contributions of this paper is the verification that this assumption is satisfied for local linear kernel in low-dimensional cases. In high-dimensional cases, this assumption may be hard to meet, and we thus propose a sample splitting algorithm for machine learning methods, which is another contribution of this paper. In brief, this algorithm separates the samples used for the estimation of $\hat{h}_{[k]}(\cdot,a)$ and the evaluation sample $X_i$ in $\hatha$, hence simplifies the required assumption. Detailed information about the sample splitting process, simplified assumption can be found in Section~\ref{subsec::sample}.

Denote $\hat{r}_i(a) = Y_i (a) - [(1-\pi) \hathts + \pi \hathcs]$ as the estimated value of $r_i(a)$. The asymptotic normality and a consistent asymptotic variance estimator of $\taupro$ is provided as in the following theorem by \cite{Liu2020Lasso}, thus justifying the Wald-type inference of the average treatment effect.
\begin{theorem}\label{Theo2}
  Suppose that $r_i(a) \in \mathcal{R}_2, \ E\{ h^2_{[k]}(X_i,a) \} < \infty, \ a = 0, 1, \ k = 1, \dots, K$. Under Assumptions~\ref{assum1}--\ref{assum4},
  $$\sqrt{n}(\taupro - \tau) \stackrel{d}\rightarrow \mathcal{N}(0,\varsigma^2_r(\pi) + \varsigma^2_{Hr}), \qquad \hat{\varsigma}^2_r(\pi) + \hat{\varsigma}^2_{Hr} \stackrel{P}\rightarrow \varsigma^2_r(\pi) + \varsigma^2_{Hr}.$$
\end{theorem}

\subsection{Optimal choice of transformed outcomes}
There are different choices of the subtractor in transformed outcomes, but not all of them improve the efficiency of the treatment effect estimator. How to achieve the optimal efficiency of the treatment effect estimator has been established as in the following theorem by \cite{Liu2020Lasso}, and we give a more detailed explanation here.
\begin{theorem}\label{Theo3}
Conditional on $B = k$, $r(a) = Y(a) - \left[\pmi E\{Y(1)|X, B=k\} + \pi \right.$\\$\left. E\{Y(0)|X, B=k\}\right]$ has the minimum variance among the sets of all transformed outcomes of the form $Y(a) - [(1-\pi)h_{[k]}(X,1)+\pi h_{[k]}(X,0)]$. In other words, the minimum variance of the treatment effect estimator is obtained when $h_{[k]}(X,a) = E\{Y(a)|X, B=k\}$.
\end{theorem}

\begin{remark}\label{remark::generalerror}
  The distance between the estimated function and the true function is often called the generalization error. As widely acknowledged, the generalization error of an estimated function can be decomposed into estimation error and approximation error \citep[e.g.,][]{Barron1994,Niyogi1996,Pinkus2012}. The estimation error is the distance between the estimated function and the optimal function that we can achieve in a restricted function space, and is determined by estimating methods. The optimal function is the projection of the outcomes on the restricted function space. The estimation error results from the fact that we are estimating functions on finite samples. A larger sample size results in a smaller the estimation error, i.e., it is more likely that the estimated function will approach the optimal function. However, the requirements on the sample sizes to achieve similar estimation errors vary from method to method, and the convergence rates of nonparametric regressions were demonstrated by \cite{Stone1980,Stone1982}. The approximation error is the distance between the true function and the optimal function. It decreases as the range of the function space increases. Theoretically, the approximation error can be eliminated if the function space is extended to close to the full space.
\end{remark}

To conduct valid and efficient inference, we need to minimize the distance between the estimation function and true function, i.e., control both the estimation error and approximation error. Assumption~\ref{assum4} imposes detailed requirements on the convergence rate of the estimating methods, which controls the estimation error to obtain valid estimates. To optimize efficiency, we need to further eliminate the approximation error, which is determined by the function space spanned by estimating methods. If the function space contains the true function (i.e., $E\{Y(a)|X, B=k\},\ a = 0, 1$ in Theorem~\ref{Theo3}), then the corresponding treatment effect estimator has optimal efficiency. However, it is often the case that we cannot know exactly which function space contains the true function, especially in high-dimensional cases. Therefore, we can only consider the rationality of function spaces for different estimating methods, such as the linear function space for linear regressions, the sieve space for artificial neural networks (ANNs), and the spline space for smoothing methods, in the context of current data, and compare their practical performances. In the following sections, we introduce commonly used estimating methods and demonstrate their empirical efficiencies through numerical simulations.

\section{Nonparametric methods for low-dimensional cases}

If a few covariates are known to be strongly correlated with the outcomes, according to domain knowledge or external data, then we can use these key covariates to make efficient statistical inferences without a heavy computational burden. Below are several models commonly used in low-dimensional settings. Note that other machine learning methods, such as neural networks and random forest, can also be applied as elaborated in Section~\ref{mlmethods}. We conduct the simulation study for machine learning methods in both low-dimensional and high-dimensional settings.

\subsection{Local linear kernel}
Although \cite{Tsiatis2008} suggested a general strategy for estimating the projection function with the flexible use of modeling methods for covariate-adjusted estimators under simple randomization, they did not give theoretical details. Several nonparametric statistical tools including kernel, spline, and orthogonal series have been proposed and widely used for estimating $h_{[k]}(X_i,a)$. Among these tools, local linear kernel has better asymptotic behavior \citep{Fan1992}. In this paper, we give a formal justification for the use of the local linear kernel smoother of $h_{[k]}(X_i,a)$ under stratified randomization. Consider the problem that for $i \in [k]$,
\begin{equation}
\operatorname{Minimize} \sumjk \left\{Y_{j}(a)-\alpha-\beta^{T}\left(X_{j}- X_i \right)\right\}^{2} K_{H}\left(X_{j}-X_i\right) \cdot \mathds{1}_{A_j = a},
\end{equation}
where $H$ is a $d \times d$ symmetric positive definite matrix depending on $n$, $K_H(u) = |H|^{-1/2} K(H^{-1/2}u)$ with $K$ being a $d$-dimensional kernel such that $\int K(u) d_u = 1$, and $| \cdot |$ denotes the determinant of a matrix. $\mathds{1}_{A_j = a}$ is an indicator function that equals $1$ if $A_j = a$ and $0$ otherwise. $H^{1/2}$ is called the bandwidth matrix. Then, $\hat{h}_{[k]}(X_i,a) = \hat{\alpha}$ is the local linear kernel smoother of $h_{[k]}(X_i,a)$. We make the following general assumptions about local linear kernel.

\begin{Assumption}\label{assum::kernel}
~

  (i) Conditional on $B_i = k$, $Y_i(a) = \ha + \nu^{1/2}(X_i)\varepsilon_i$, where $\nu(x) = \Var(Y|X = x)>0$ is continuous, $E\{ h^2_{[k]}(X_i,a) \} < \infty, \ a=0,1, \ k=1,\dots,K$, and the probability density function of $X_i$'s has a compact support set on $R^d$. $\varepsilon_i$'s are mutually i.i.d. random variables with zero mean and unit variance and are independent of $X_i$. All second-order derivatives of $h_{[k]}(\cdot,a)$ are continuous.

  (ii) The kernel $K$ is a compactly supported, bounded kernel such that $\int uu^\T K(u) du = \mu_2(K)I$, where $\mu_2(K) \neq 0$ is a scalar and $I$ is the $d \times d$ identity matrix. Additionlly, all odd-order moments of $K$ vanish, i.e., $\int u_1^{l1}\cdots u_d^{l_d} K(u) du = 0$ for all nonnegative integers $l_1,\dots,\l_d$ such that their sum is odd.

  (iii) $n^{-1}|H|^{-1}$ and each entry of $H$ tend to zero as $n \to \infty$, with $H$ remaining symmetric and positive definite. Moreover, there is a fixed constant $L$ such that the condition number of $H$ is at most $L$ for all $n$.

\end{Assumption}

\vspace{0.1cm}
Then based on Theorem~\ref{Theo2}, we establish the following theorem.
\begin{theorem}
\label{Theo5}
  Under Assumptions~\ref{assum1}--\ref{assum3} and \ref{assum::kernel}, for $k \in 1,\dots,K$ and $a = 0,1$, $\hatha$ satisfy Assumption~\ref{assum4}. Therefore, the adjusted treatment effect estimators obtained using local linear kernel allow for valid statistical inference.
\end{theorem}
\vspace{0.1cm}

\begin{remark}
The detailed proof for local linear kernel can be found in Section 3 of the Supplementary Materials. \cite{gelman2019high} stated that in regression discontinuity analysis, higher-order polynomials are no better than local linear or quadratic polynomials mainly for three reasons, two of which also apply to our situations. First, if we rewrite the polynomial regression estimator as the weighted average of the outcomes, we will find that the higher-order terms actually make no contribution to our estimated function of interest, which is the first component of the estimator. Second, the use of higher-order polynomial regression requires extra knowledge of the existence of higher-order derivatives of the function to be estimated, and there is no universal method with which to make a good determination yet. Additionally, the superiority of local linear kernel is also demonstrated through a simulation study in Section~\ref{sec::simstudy}.
\end{remark}

\subsection{Spline smoothing}
As a nonparametric method competing with kernel methods, spline smoothing has commonly been investigated in the literature. Unlike the local linear weighted regression idea of the kernel, spline smoothing considers adding penalties when minimizing the sum of squares, to avoid over-fitting. One of the most frequently adopted penalties is the integral over the second-order derivative of the estimated function, which corresponds to the cubic spline. Additionally, despite the different forms and concepts of kernel and spline, \cite{Silverman1984} showed the asymptotic equivalence of spline smoothing and a kernel method with a bandwidth depending on the local density of design points. Our simulation results suggest the similar performance of spline smoothing and kernel methods under certain scenarios and the superiority of local linear kernel in certain situations. Thus, we do not explore the theoretical justification of spline smoothing-adjusted estimators in this paper.

\section{Machine-learning methods for high-dimensional cases}\label{mlmethods}
In the era of big data, a massive amount of information can be collected, and it is difficult for us to artificially determine the most relevant covariates. Additionally, the interactions between covariates are difficult to approach by simple forms. Therefore, traditional statistical methods may not be suitable in this situation, and high-dimensional methods such as the lasso and other machine learning methods need to intervene.

In general, high-dimensional methods commonly used nowadays perform well in model prediction, which is what they were mainly designed for, but these methods are lacking in estimation efficiency and valid statistical inference. The theoretical properties of estimators adjusted using high-dimensional methods such as tree-based methods and neural networks are difficult to verify and remain to be uncovered, especially under stratified randomization. However, the ability of the high-dimensional methods to capture nonlinear features and interactions is evident. In this paper, we mainly report a comparison of the estimators adjusted using different high-dimensional techniques in the numerical study and present the pros and cons of the estimators.

\subsection{Penalized regression}
In the presence of high-dimensional covariates, linear regression is prone to problems of multi-collinearity or overfitting. In practice, penalized regressions are usually adopted to solve these problems. When the exact or weak sparsity assumption of the population projection coefficients is reasonable, lasso can be adopted, and the satisfiability of the lasso-adjusted estimators for Assumption~\ref{assum4} was proved by \cite{Liu2020Lasso}. Fundamental work on the concentration inequality and restricted eigenvalue under stratified randomization in the work of \cite{Liu2020Lasso} can be used for other penalized regressions, such as adaptive-lasso \citep{Zou2006, HuangZhang2008}, Ridge regression \citep{Hoerl1970}, and elastic net regression \citep{Zou2005}. Moreover, although the debiased-lasso \citep{Zhang2014debiased} usually has a performance similar to that of lasso, it has a higher computational cost, and it is thus not recommended here. When there are many weak predictors, Ridge regression can be chosen.

\subsection{General machine learning methods}\label{subsec::ml}

With the abundance of available data and the increase of model complexity, we need the help of computers to perform data-intensive and complex operations in addition to model training and estimation, i.e., machine learning. Nowadays, there are many machine learning methods available to researchers, among which the two types of method commonly used in medical research are tree-based methods and neural networks \citep{Garg2021}. Both types of method can handle data with high dimensionality and complex interactions through stepwise deconstruction. However, as has been pointed out, no single method can solve problems in a one-size-fits-all manner \citep[e.g.,][]{Peel2010,Pedersen2020,De2021}. In practice, we should consider the type of problem to be solved, the structure of covariates, and the parameters of interest in selecting the most appropriate method. The good approximability of machine learning methods effectively reduces the approximation error.

Tree-based methods hierarchically partition the covariate space, recursively dividing the entire space into small regions. They can handle categorical and ordinal covariates in a simple way and automatically select covariates in steps and reduce model complexity. However, the single-tree structure tends to have insufficient prediction accuracy. The prediction performance has thus been improved using ensemble trees, commonly through bagging and boosting. Random forest, proposed by \cite{Breiman2001}, is a substantial modification of bagging \citep{Breiman1996} in ensemble learning that constructs de-correlated trees on different bootstrap samples of the data and averages the results. Random forest is widely used in randomized experiments. For example, \cite{Wu2018} proposed a ``leave-one-out potential outcomes" estimator by imputing potential outcomes using random forest, and \cite{wager2018} estimated and inferred heterogeneous treatment effects using random forest. Boosting is a sequential process that continuously trains weak classifiers or regressors and adjusts the weights of samples and classifiers or regressors in each iteration, to reduce the prediction error. The tree-based methods that are widely used in practice in combination with boosting are the gradient boosting decision tree \citep{Friedman2002}, Adaboost \citep{Freund1997adaboost} and XGboost \citep{Chen2015xgboost}.

Benefitting from the rapid development of computer technology, ANNs are appealing machine learning methods that can approach a wide variety of (nonlinear) functions, especially in the case of high-dimensional data \citep[e.g.,][]{White1992,Yarotsky2018,Schmidt2020}. The willingness to use ANNs has been demonstrated in different areas of research, including pattern recognition, decision-making, and pharmaceutical research. In recent years, ANNs have also been introduced to improve the efficiency of estimating treatment effects. \cite{Farrell2021} used ANNs to estimate nuisance functions in the estimation procedure of treatment effects, based on efficient influence functions. \cite{Chen2020} considered a more general framework of treatment effects with propensity score functions estimated using ANNs.

\subsection{Sample splitting}\label{subsec::sample}

Machine learning methods can well estimate projection functions and make better predictions than traditional methods, but the fitted functions may have plenty of parameters and be complex in form. Additionally, substantive biases induced by machine learning methods are inevitable, and the restrictions on the estimation error in Assumption~\ref{assum4} are thus difficult to verify and may not be satisfied for various machine learning methods. In this case, we can use sample splitting for the separate estimation of the projection function and treatment effect, which relaxes the consistency conditions and thus makes most machine learning methods feasible \citep{Chernozhukov2018}.

Sample splitting is a common technique in statistics. This technique usually involves dividing data into two parts, one part for the inference of parameters or functions of interest and the other part for validation or estimating nuisance parameters \citep{Picard1990}. Through sample splitting, we gain accuracy and robustness of the inference at the cost of prediction efficiency owing to the reduction of the sample size \citep{Rinaldo2019}. To regain full efficiency, \cite{Chernozhukov2018} proposed a ``cross-fitting'' process, which includes dividing data evenly into $M$ parts, estimating the parameter of interest and its variance on each inference-estimation data pair, and averaging them to obtain the final estimator. As the sample size increases, the estimators obtained from inference-estimation data pairs become asymptotically independent, and their variances can thus be aggregated for inference. The adoption of additional orthogonalization to reduce the regularization bias is known as the double/debiased machine learning method, which was proposed by \cite{Chernozhukov2018}. This method has gained prevalence in research on high-dimensional statistical inference \citep[e.g.,][]{Kallus2019,Bodory2020}.

Moreover, the asymptotic properties of the methods described in Section~\ref{subsec::ml} and the double/debiased machine learning method are deduced under the assumption that the treatment assignments and thus the outcomes are independent. However, under stratified randomization, there may be correlations between outcomes and between treatment assignment indicators within each stratum because of the treatment assignment procedure, which may lead to the invalidity of the aforementioned statistical inference. This motivates us to combine the form of the variance estimator in Theorem~\ref{Theo2} with the sample splitting technique, where we expect to obtain a valid statistical inference using machine learning methods under stratified randomization. The proposed algorithm is described in Algorithm~\ref{alg1}.

\begin{algorithm}
	\renewcommand{\algorithmicrequire}{\textbf{Input:}}
	\renewcommand{\algorithmicensure}{\textbf{Output:}}
	\caption{Sample splitting algorithm for estimating the average treatment effect under stratified randomization.}
	\label{alg1}
	\begin{algorithmic}[1]
\vskip 1mm
		\State Take an (almost) evenly $M$-fold random partition $(I_m)_{m=1}^M$ of the observed indices $\{1,\dots,n\}$, for $m = 1,\dots, M-1$, the size of fold $m$ is $\lfloor n/M \rfloor$, and the size of the last fold is $n-(M-1)\lfloor n/M \rfloor$. Define the complementary set of $I_m$ as $I_m^c = \{i:i \in \{1,\dots,n\}, \ i \notin I_m\}$
		\State  For each fold $m \in \{1,\dots,M\}$, construct an estimator $\hat{h}_{[k]m}(\cdot,a)$ of $h_{[k]}(\cdot,a)$ based on data indexed by $I_m^c$
		\State Plug in $\hat{h}_{[k]m}(\cdot,a)$ to the oracle estimator and use data indexed by $I_m$ for the inference of the average treatment effect. That is, for each $m \in \{1,\dots,M\}$, obtain a treatment effect estimator $\hat{\tau}_{m}$ as in Equation~\eqref{eqn::emp} and a variance estimator $\hat{\sigma}^2_{m}$ by Theorem~\ref{Theo2}
		\State Aggregate the treatment effect estimators:
  \begin{equation}\label{eqn::tauds}
    \tauss = \frac{1}{M}\sum_{m=1}^M \hat{\tau}_{m}. \nonumber
  \end{equation}
  The variance estimator of $\sqrt{n}\tauss$ is
  \begin{equation}\label{eqn::vards}
    \hat{\sigma}^2_{ss} = \frac{1}{M} \sum_{m=1}^M \hat{\sigma}^2_{m}. \nonumber
  \end{equation}

	\end{algorithmic}
\end{algorithm}

Indeed, this algorithm incorporates the sample splitting technique into the general framework of inference under stratified randomization. Using this algorithm, Assumption~\ref{assum4} on the projection function can reduce to the following second moment convergence assumption.
\begin{Assumption}\label{assum::ss}
Denote the $M$-fold random partition of the observed indices $\{1,\dots,n\}$ as $(I_m)_{m=1}^M$, for fold $m = 1,\dots, M,$
\begin{equation}
\label{eqn::goalss}
  E \Big[\big\{   \hat{h}_{[k]m}(\tilde X_i,a) - h_{[k]}(\tilde X_i,a)  \big\}^2 \mid B_i = k, \{Y_j, X_j, A_j, B_j\}_{j \in I_{m}^c} \Big] = o_P(1),
\end{equation}
where $\tilde X_i$ has the same distribution as $\{X_i \mid B_i = k\}$ and is independent of $\{Y_j, X_j, A_j, B_j\}_{j \in I_{m}^c}$, and $\hat{h}_{[k]m}(\cdot,a)$ is the projection function estimated by samples not in fold $m$.
\end{Assumption}
\vspace{0.1cm}

The above assumption can be satisfied by many machine learning methods under certain assumptions \citep{Chernozhukov2018}, and we establish the following theorem.
\vspace{0.1cm}
\begin{theorem}\label{theo::ss}
  Suppose that $r_{i}(a) \in \mathcal{R}_2, \ E\{ h^2_{[k]}(X_i,a) \} < \infty, \ a = 0, 1, i \in [k] \cap I_m , \ k = 1, \dots, K, \ m = 1, \dots, M$. Then under Assumptions~\ref{assum1}--\ref{assum3} and Assumption~\ref{assum::ss},
  $$\sqrt{n}(\hat{\tau}_{ss} - \tau) \stackrel{d}\rightarrow \mathcal{N}(0,\varsigma^2_r(\pi) + \varsigma^2_{Hr}), \qquad \hat{\sigma}^2_{ss} \stackrel{P}\rightarrow \varsigma^2_r(\pi) + \varsigma^2_{Hr},$$
  where $\hat{\tau}_{ss}$ and $\hat{\sigma}^2_{ss}$ are defined in Algorithm~\ref{alg1}.
\end{theorem}

From the subsequent simulation results (Table~\ref{tab::equalsimhighds1000}), we can see that all the treatment effect estimators obtained using Algorithm~\ref{alg1} enjoy unbiasedness and validity.

\section{Simulation study}\label{sec::simstudy}

In this section, we examine the empirical performance of estimators with different estimating methods for $\hax, \ a = 0, 1$. We consider four low-dimensional data-generating models and four corresponding high-dimensional data-generating models with continuous outcomes. For $a \in \{0,1\}$ and  $1 \leq i \leq n$, the potential outcomes are generated according to
$$Y_i(a) = g_a(X_i) + \sigma_a \varepsilon_{a,i},$$
where $X_i, \ g_a(X_i), \ i = 1, \dots, n$, are specified below. In each model, $(X_i, \varepsilon_{0,i}, \varepsilon_{1,i}), \ 1 \leq i \leq n$ are i.i.d., and we set $\sigma_0 = 1, \ \sigma_1 = 3$. Both $\varepsilon_{0,i}$ and $\varepsilon_{1,i}$ follow the standard normal distribution. For high-dimensional data-generating models with $p$ covariates, there are few covariates that truly correlate with outcomes, and we generate additional independent covariates to approach reality.

Here, we present the simulation results of the estimators under simple randomization, stratified block randomization, and minimization. The sample size $n$ is $1000$. The block size used in stratified block randomization is $6$. A biased-coin probability of $0.75$ and equal weights are used in minimization. The bias, standard deviation (SD) of the treatment effect estimators, standard error (SE) estimators, and empirical coverage probability (CP) of the $95\%$ confidence interval are computed using $2000$ replications.

\subsection{Low-dimensional data-generating models}

Model 1:
$$\begin{aligned}
g_0(X_i) &=\mu_0 + \sum\limits_{j=1}^4 \beta_{0j} X_{ij}, \\
g_1(X_i) &=\mu_1 + \sum\limits_{j=1}^4 \beta_{1j} X_{ij},
\end{aligned}$$
with $\mu_0 = 1, \ \mu_1 = 4, \ \beta_0^{\textnormal{T}} =  (75, 35, 125, 80)$, and $\beta_1^{\textnormal{T}} =  (100, 80, 60, 40)$. $X_i$ is a four-dimensional vector, $X_{i1} \sim \textup{Beta}(3,4)$, $X_{i2} \sim \textup{Unif}[-2,2]$, $X_{i3}$ takes values in $\{-1, 1\}$ with equal probability, $X_{i4}$ takes values in $\{3, 5\}$ with probability $0.6, 0.4$, and they are independent of each other. The variable used for randomization is an additional variable that takes a value in $\{1, 2, 3, 4\}$ with probability $0.2, 0.3, 0.3, 0.2$ and is independent of $X_{ij}$. Model 1 considers the regular linear model as a baseline for other nonlinear models.

Model 2:
$$\begin{aligned}
g_0(X_i) &=\mu_0 + \beta_{01}\log(X_{i1}+1) + \beta_{02}X_{i1}^2 + \beta_{03}\exp(X_{i2}) + \beta_{04}/(X_{i2}+3), \\
g_1(X_i) &=\mu_1 + \beta_{11}\exp(X_{i1}+2) + \beta_{12}/(X_{i1}+1) + \beta_{13}X_{i2}^2,
\end{aligned}$$
with $\mu_0 = -3,\  \mu_1 = 0,\  \beta_0^{\textnormal{T}} = (10, 24, 15, 20)$, and $\beta_1^{\textnormal{T}} =  (20, 17, 10)$. $X_i$ is a two-dimensional vector, $X_{i1} \sim \textup{Beta}(3,4)$, $X_{i2} \sim \textup{Unif}[-2,2]$, and they are independent of each other. The variable used for randomization is an additional variable that takes a value in $\{1, 2, 3, 4\}$ with probability $0.2, 0.3, 0.3, 0.2$ and is independent of $X_{ij}$. Model 2 is an additive but nonlinear model of covariates.

Model 3:
$$\begin{aligned}
g_0(X_i) &=\mu_0 + \beta_{01}X_{i1}X_{i2}/(X_{i1}+X_{i2}+2)+\beta_{02}X_{i1}^2(X_{i2}+X_{i3}), \\
g_1(X_i) &=\mu_1 + \beta_{11}(X_{i2}+X_{i4}) + \beta_{12}X_{i2}^2/\exp(X_{i1}+2),
\end{aligned}$$
with $\mu_0 = 5, \ \mu_1 = 2, \ \beta_0^{\textnormal{T}} = (42, 83)$, and $\beta_1^{\textnormal{T}} = (30, 75)$. $X_i$ is a four-dimensional vector, $X_{i1} \sim \textup{Beta}(3,4)$, $X_{i2} \sim \textup{Unif}[-2,2]$, $X_{i3} \sim \mathcal{N}(0,1)$, $X_{i4} \sim \textup{Unif}[0,2]$, and they are independent of each other. The variable used for randomization is an additional variable that takes a value in $\{1, 2\}$ with probability $0.4, 0.6$ and is independent of $X_{ij}$. Model 3 is a nonlinear model including interaction terms of covariates, hence is more complex.

Model 4:
$$\begin{aligned}
g_0(X_i) &= \mu_0 + (\beta_{01}X_{i1} + \beta_{02}X_{i2})S + \beta_{03}\log(X_{i1}+1)\mathds{1}_{S_i = 1}, \\
g_1(X_i) &= \mu_1 + (\beta_{11}X_{i1} + \beta_{12}X_{i2})S + \beta_{13}\exp(X_{i2})\mathds{1}_{S_i = -1},
\end{aligned}$$
with $\mu_0 = 5, \ \mu_1 = 5, \ \beta_0^{\textnormal{T}} = (20, 30, 50)$, and $\beta_1^{\textnormal{T}} = (20, 30, 65)$. $X_i$ is a two-dimensional vector, $X_{i1} \sim \textup{Beta}(3,4)$, $X_{i2} \sim \textup{Unif}[-2,2]$, and they are independent of each other. $\mathds{1}_{S_i = 1}$ is an indicator function that equals $1$ if $S_i = 1$ and $0$ otherwise. $\mathds{1}_{S_i = -1}$ is defined likewise. $S_i$ is the variable used for randomization, it is an additional variable that takes a value in $\{1, -1\}$ with equal probability and is independent of $X_{ij}$. Model 4 further takes the interaction between covariates and stratum into consideration.

\subsection{Low-dimensional simulation results}
For low-dimensional data-generating models, we apply linear regression, local linear kernel (kernel), natural spline (nspline), neural networks (nnet), and random forest (rf) to approach $\hatha$. In the result tables, $\hat{\tau}$ denotes the stratum-common estimators ($\hat h_{[k]}(X_i,a)$ are the same for all strata $k, \ k = 1, \dots, K$) and $\tilde{\tau}$ denotes the stratum-specific estimators ($\hat h_{[k]}(X_i,a)$ can be different in each stratum).

From Table~\ref{tab::equalsim1000}, we see that there is little difference among the treatment effect estimators when different randomization methods are used. Under all considered scenarios, the treatment effect estimators obtained using different covariate-adjustment methods have small biases. When the data-generating model is linear (Model 1), the linear regression-adjusted estimators have the optimal efficiency, whereas $\tauker$ and $\tauns$ have the same efficiency and are followed by $\taunnet$ and then $\taurf$. When the data-generating model is nonlinear (Models 2, 3 and 4), all $\taulin$s always have relatively large standard error. Therefore, there can be a loss of efficiency when stubbornly using linear models for covariate adjustment, especially when there is strong evidence of a nonlinear relationship between covariates and outcomes. In this case, we should consider using other methods for covariate adjustment.

For each data-generating model, the fitting performance differs among the adjusting methods. For example, in Model 2, $\tauker$ and $\tauns$ have the minimum standard deviations, followed by $\taunnet$ and then $\taurf$. However, in the cases of Models 3 and 4, $\taurf$ has the largest standard deviations among the treatment effect estimators adjusted using nonlinear methods. In Model 3, the standard deviation of $\taurf$ is even greater than that of $\taulin$. A comparison between stratum-common and stratum-specific estimators reveals that the stratum-specific estimators have smaller standard deviations than the corresponding stratum-common estimators only if the data-generating model is stratum-specific (Model 4).

From the perspective of statistical inference, all treatment effect estimators have the desired coverage probability, except the estimator adjusted using random forest. Particularly, in Model 4, the coverage probabilities of $\taurf$ are below 0.9, showing the unsatisfiability of the proposed assumptions. This inspires us to further use other statistical techniques to reach valid statistical inferences. Similar conclusions under unequal allocation ($\pi = 2/3$) are shown in Table~\ref{tab::unequalsim1000}.

\begin{table}
	\centering
	\caption{Simulated biases, standard deviations, standard errors, and coverage probabilities for different estimators and randomization methods under equal allocation ($\pi=1/2$) and low-dim-\\ensional data-generating models.}\label{tab::equalsim1000}
	\vskip 1.4mm
	\begin{threeparttable}
		\setlength{\tabcolsep}{1.4pt}
\resizebox{0.8\textwidth}{90mm}{
		\begin{tabular}{llccccccccccccccc}
		\cline{1-17}
		&  & \makebox[0.025\textwidth][c]{} &
        \multicolumn{4}{c}{Complete Rand.} &
        \makebox[0.025\textwidth][c]{} &
        \multicolumn{4}{c}{Stratified Block Rand.} &
        \makebox[0.025\textwidth][c]{} &
        \multicolumn{4}{c}{Minimization} \\ \cline{3-17}
		Model &
		Estimator & &
		\multicolumn{1}{c}{Bias} &
		\multicolumn{1}{c}{SD} &
		\multicolumn{1}{c}{SE} &
        \multicolumn{1}{c}{CP} & &
		\multicolumn{1}{c}{Bias} &
		\multicolumn{1}{c}{SD} &
		\multicolumn{1}{c}{SE} &
        \multicolumn{1}{c}{CP} & &
		\multicolumn{1}{c}{Bias} &
		\multicolumn{1}{c}{SD} &
		\multicolumn{1}{c}{SE} &
        \multicolumn{1}{c}{CP}  \\
  \hline
  1 & $\taulin$ &  & 0.02 & 3.00 & 2.91 & 0.95 &  & -0.08 & 2.89 & 2.91 & 0.95 &  & -0.03 & 2.98 & 2.91 & 0.95 \\
  & $\ttaulin$ &  & 0.02 & 3.00 & 2.91 & 0.95 &  & -0.08 & 2.89 & 2.91 & 0.95 &  & -0.03 & 2.98 & 2.91 & 0.95 \\
  & $\tauker$ &  & 0.02 & 3.00 & 2.91 & 0.95 &  & -0.08 & 2.89 & 2.91 & 0.95 &  & -0.03 & 2.98 & 2.91 & 0.95 \\
  & $\ttauker$ &  & 0.02 & 3.00 & 2.91 & 0.95 &  & -0.08 & 2.89 & 2.91 & 0.95 &  & -0.03 & 2.98 & 2.91 & 0.95 \\
  & $\tauns$ &  & 0.02 & 3.00 & 2.91 & 0.95 &  & -0.07 & 2.89 & 2.91 & 0.95 &  & -0.03 & 2.98 & 2.91 & 0.95 \\
  & $\ttauns$ &  & 0.02 & 3.00 & 2.91 & 0.95 &  & -0.07 & 2.89 & 2.91 & 0.95 &  & -0.03 & 2.98 & 2.91 & 0.95 \\
  & $\taunnet$ &  & 0.02 & 3.01 & 2.91 & 0.95 &  & -0.07 & 2.89 & 2.91 & 0.95 &  & -0.03 & 2.99 & 2.91 & 0.95 \\
  & $\ttaunnet$ &  & 0.00 & 3.01 & 2.93 & 0.95 &  & -0.08 & 2.90 & 2.92 & 0.95 &  & -0.04 & 3.00 & 2.92 & 0.95 \\
  & $\taurf$ &  & 0.00 & 4.14 & 3.97 & 0.94 &  & -0.12 & 4.14 & 3.97 & 0.94 &  & -0.07 & 4.20 & 3.97 & 0.94 \\
  \vspace{0.25cm}
  & $\ttaurf$ &  & 0.01 & 4.37 & 4.05 & 0.92 &  & -0.09 & 4.40 & 4.04 & 0.93 &  & -0.06 & 4.46 & 4.04 & 0.92 \\
  2 & $\taulin$ &  & -0.02 & 1.49 & 1.52 & 0.95 &  & -0.03 & 1.51 & 1.52 & 0.95 &  & 0.04 & 1.52 & 1.52 & 0.95 \\
  & $\ttaulin$ &  & -0.03 & 1.49 & 1.52 & 0.95 &  & -0.04 & 1.51 & 1.52 & 0.95 &  & 0.03 & 1.52 & 1.52 & 0.96 \\
  & $\tauker$ &  & 0.02 & 1.27 & 1.28 & 0.95 &  & 0.00 & 1.27 & 1.28 & 0.95 &  & 0.06 & 1.29 & 1.28 & 0.95 \\
  & $\ttauker$ &  & 0.07 & 1.27 & 1.28 & 0.95 &  & 0.05 & 1.27 & 1.28 & 0.95 &  & 0.11 & 1.30 & 1.28 & 0.95 \\
  & $\tauns$ &  & 0.00 & 1.27 & 1.28 & 0.95 &  & -0.02 & 1.26 & 1.28 & 0.95 &  & 0.04 & 1.29 & 1.28 & 0.95 \\
  & $\ttauns$ &  & 0.00 & 1.27 & 1.28 & 0.95 &  & -0.02 & 1.26 & 1.28 & 0.95 &  & 0.03 & 1.29 & 1.28 & 0.95 \\
  & $\taunnet$ &  & -0.02 & 1.29 & 1.30 & 0.95 &  & -0.03 & 1.29 & 1.30 & 0.95 &  & 0.02 & 1.31 & 1.30 & 0.95 \\
  & $\ttaunnet$ &  & -0.06 & 1.35 & 1.36 & 0.95 &  & -0.08 & 1.36 & 1.36 & 0.95 &  & -0.01 & 1.37 & 1.36 & 0.95 \\
  & $\taurf$ &  & 0.00 & 1.28 & 1.27 & 0.95 &  & -0.02 & 1.29 & 1.27 & 0.94 &  & 0.04 & 1.31 & 1.27 & 0.95 \\
  \vspace{0.25cm}
  & $\ttaurf$ &  & 0.04 & 1.32 & 1.27 & 0.94 &  & 0.01 & 1.32 & 1.27 & 0.94 &  & 0.08 & 1.35 & 1.27 & 0.94 \\
  3 & $\taulin$ &  & 0.00 & 1.39 & 1.41 & 0.96 &  & -0.06 & 1.39 & 1.41 & 0.96 &  & -0.03 & 1.37 & 1.41 & 0.95 \\
  & $\ttaulin$ &  & -0.02 & 1.39 & 1.40 & 0.96 &  & -0.08 & 1.40 & 1.40 & 0.95 &  & -0.04 & 1.37 & 1.40 & 0.95 \\
  & $\tauker$ &  & 0.08 & 1.21 & 1.19 & 0.94 &  & 0.04 & 1.18 & 1.19 & 0.95 &  & 0.05 & 1.19 & 1.19 & 0.95 \\
  & $\ttauker$ &  & 0.13 & 1.21 & 1.19 & 0.94 &  & 0.10 & 1.19 & 1.19 & 0.95 &  & 0.10 & 1.20 & 1.19 & 0.94 \\
  & $\tauns$ &  & 0.01 & 1.39 & 1.38 & 0.95 &  & -0.05 & 1.41 & 1.38 & 0.95 &  & -0.02 & 1.38 & 1.38 & 0.95 \\
  & $\ttauns$ &  & 0.00 & 1.40 & 1.38 & 0.95 &  & -0.05 & 1.40 & 1.38 & 0.94 &  & -0.03 & 1.40 & 1.38 & 0.94 \\
  & $\taunnet$ &  & 0.02 & 1.29 & 1.29 & 0.95 &  & -0.05 & 1.29 & 1.29 & 0.95 &  & -0.03 & 1.28 & 1.29 & 0.95 \\
  & $\ttaunnet$ &  & -0.03 & 1.31 & 1.31 & 0.95 &  & -0.07 & 1.31 & 1.31 & 0.95 &  & -0.05 & 1.31 & 1.31 & 0.95 \\
  & $\taurf$ &  & 0.00 & 1.40 & 1.22 & 0.92 &  & -0.03 & 1.39 & 1.22 & 0.91 &  & -0.02 & 1.40 & 1.22 & 0.91 \\
  \vspace{0.25cm}
  & $\ttaurf$ &  & -0.01 & 1.48 & 1.26 & 0.91 &  & -0.04 & 1.48 & 1.26 & 0.90 &  & -0.03 & 1.49 & 1.26 & 0.90 \\
  4 & $\taulin$ &  & -0.11 & 3.67 & 3.71 & 0.95 &  & 0.03 & 3.82 & 3.71 & 0.95 &  & -0.06 & 3.76 & 3.71 & 0.95 \\
  & $\ttaulin$ &  & -0.12 & 3.61 & 3.67 & 0.96 &  & 0.01 & 3.8 & 3.67 & 0.94 &  & -0.10 & 3.72 & 3.67 & 0.94 \\
  & $\tauker$ &  & -0.05 & 3.63 & 3.57 & 0.94 &  & 0.11 & 3.73 & 3.57 & 0.94 &  & 0.00 & 3.68 & 3.56 & 0.94 \\
  & $\ttauker$ &  & -0.05 & 3.45 & 3.47 & 0.95 &  & 0.11 & 3.56 & 3.47 & 0.95 &  & -0.02 & 3.53 & 3.47 & 0.94 \\
  & $\tauns$ &  & -0.11 & 3.64 & 3.59 & 0.95 &  & 0.10 & 3.72 & 3.59 & 0.94 &  & -0.03 & 3.66 & 3.59 & 0.94 \\
  & $\ttauns$ &  & -0.07 & 3.45 & 3.47 & 0.95 &  & 0.09 & 3.56 & 3.47 & 0.95 &  & -0.03 & 3.53 & 3.47 & 0.95 \\
  & $\taunnet$ &  & -0.11 & 3.59 & 3.61 & 0.95 &  & 0.08 & 3.70 & 3.61 & 0.95 &  & -0.05 & 3.65 & 3.61 & 0.94 \\
  & $\ttaunnet$ &  & -0.08 & 3.45 & 3.47 & 0.95 &  & 0.08 & 3.56 & 3.48 & 0.95 &  & -0.04 & 3.53 & 3.47 & 0.94 \\
  & $\taurf$ &  & -0.14 & 3.69 & 3.05 & 0.89 &  & 0.05 & 3.79 & 3.06 & 0.89 &  & -0.05 & 3.75 & 3.05 & 0.88 \\
  & $\ttaurf$ &  & -0.09 & 3.47 & 3.44 & 0.95 &  & 0.08 & 3.58 & 3.44 & 0.94 &  & -0.05 & 3.54 & 3.44 & 0.94 \\
   \hline
		\end{tabular}}
\vskip 1.4mm
\begin{tablenotes}
\item Abbreviations: SD, standard deviation; SE, standard error; CP, coverage probability; Rand.: ran
    domization; nspline: natural spline; nnet: neural network; rf: random forest.
\end{tablenotes}
\end{threeparttable}
\end{table}

\begin{table}
	\centering
	\caption{Simulated biases, standard deviations, standard errors, and coverage probabilities for different estimators and randomization methods under unequal allocation ($\pi=2/3$) and low-dim-\\ensional data-generating models.}\label{tab::unequalsim1000}
	\vskip 1.4mm
	\begin{threeparttable}
		\setlength{\tabcolsep}{2pt}
\resizebox{0.8\textwidth}{90mm}{
		\begin{tabular}{llccccccccccccccc}
		\cline{1-17}
		&  & \makebox[0.025\textwidth][c]{} &
        \multicolumn{4}{c}{Complete Rand.} &
        \makebox[0.025\textwidth][c]{} &
        \multicolumn{4}{c}{Stratified Block Rand.} &
        \makebox[0.025\textwidth][c]{} &
        \multicolumn{4}{c}{Minimization} \\ \cline{3-17}
		Model &
		Estimator & &
		\multicolumn{1}{c}{Bias} &
		\multicolumn{1}{c}{SD} &
		\multicolumn{1}{c}{SE} &
        \multicolumn{1}{c}{CP} & &
		\multicolumn{1}{c}{Bias} &
		\multicolumn{1}{c}{SD} &
		\multicolumn{1}{c}{SE} &
        \multicolumn{1}{c}{CP} & &
		\multicolumn{1}{c}{Bias} &
		\multicolumn{1}{c}{SD} &
		\multicolumn{1}{c}{SE} &
        \multicolumn{1}{c}{CP}  \\
    \hline
  1 & $\taulin$ &  & 0.01 & 3.00 & 2.91 & 0.94 &  & 0.03 & 2.93 & 2.91 & 0.95 &  & 0.08 & 2.86 & 2.91 & 0.96 \\
  & $\ttaulin$ &  & 0.00 & 2.99 & 2.91 & 0.94 &  & 0.03 & 2.93 & 2.91 & 0.95 &  & 0.08 & 2.86 & 2.91 & 0.96 \\
  & $\tauker$ &  & 0.01 & 3.00 & 2.91 & 0.94 &  & 0.03 & 2.93 & 2.91 & 0.95 &  & 0.08 & 2.86 & 2.91 & 0.96 \\
  & $\ttauker$ &  & 0.00 & 2.99 & 2.91 & 0.94 &  & 0.03 & 2.93 & 2.91 & 0.95 &  & 0.08 & 2.86 & 2.91 & 0.96 \\
  & $\tauns$ &  & 0.01 & 3.00 & 2.91 & 0.94 &  & 0.03 & 2.93 & 2.91 & 0.95 &  & 0.08 & 2.86 & 2.91 & 0.96 \\
  & $\ttauns$ &  & 0.00 & 2.99 & 2.91 & 0.94 &  & 0.03 & 2.93 & 2.91 & 0.95 &  & 0.08 & 2.86 & 2.91 & 0.96 \\
  & $\taunnet$ &  & 0.00 & 3.00 & 2.91 & 0.94 &  & 0.03 & 2.94 & 2.91 & 0.95 &  & 0.08 & 2.86 & 2.91 & 0.96 \\
  & $\ttaunnet$ &  & 0.00 & 3.02 & 2.93 & 0.94 &  & 0.04 & 2.96 & 2.93 & 0.95 &  & 0.09 & 2.89 & 2.93 & 0.95 \\
  & $\taurf$ &  & -0.05 & 4.34 & 4.03 & 0.94 &  & 0.06 & 4.28 & 4.02 & 0.93 &  & 0.13 & 4.24 & 4.02 & 0.94 \\
  \vspace{0.25cm}
  & $\ttaurf$ &  & 0.00 & 4.73 & 4.12 & 0.91 &  & 0.10 & 4.69 & 4.11 & 0.91 &  & 0.17 & 4.64 & 4.10 & 0.92 \\
  2 & $\taulin$ &  & 0.04 & 1.61 & 1.55 & 0.94 &  & 0.01 & 1.56 & 1.55 & 0.96 &  & 0.02 & 1.51 & 1.55 & 0.96 \\
  & $\ttaulin$ &  & 0.09 & 1.62 & 1.55 & 0.94 &  & 0.06 & 1.56 & 1.55 & 0.96 &  & 0.07 & 1.51 & 1.54 & 0.96 \\
  & $\tauker$ &  & 0.02 & 1.31 & 1.28 & 0.94 &  & 0.00 & 1.30 & 1.28 & 0.95 &  & 0.01 & 1.27 & 1.28 & 0.95 \\
  & $\ttauker$ &  & 0.07 & 1.31 & 1.28 & 0.94 &  & 0.05 & 1.30 & 1.28 & 0.95 &  & 0.06 & 1.27 & 1.28 & 0.95 \\
  & $\tauns$ &  & 0.02 & 1.31 & 1.28 & 0.94 &  & 0.00 & 1.30 & 1.28 & 0.95 &  & 0.01 & 1.27 & 1.28 & 0.95 \\
  & $\ttauns$ &  & 0.02 & 1.31 & 1.28 & 0.94 &  & 0.00 & 1.30 & 1.28 & 0.95 &  & 0.01 & 1.27 & 1.28 & 0.95 \\
  & $\taunnet$ &  & 0.02 & 1.33 & 1.29 & 0.94 &  & 0.00 & 1.31 & 1.29 & 0.95 &  & 0.01 & 1.29 & 1.29 & 0.95 \\
  & $\ttaunnet$ &  & 0.07 & 1.43 & 1.37 & 0.94 &  & 0.04 & 1.39 & 1.36 & 0.95 &  & 0.04 & 1.35 & 1.36 & 0.95 \\
  & $\taurf$ &  & 0.01 & 1.33 & 1.27 & 0.94 &  & 0.00 & 1.31 & 1.27 & 0.94 &  & 0.00 & 1.29 & 1.27 & 0.95 \\
  \vspace{0.25cm}
  & $\ttaurf$ &  & 0.06 & 1.39 & 1.28 & 0.93 &  & 0.04 & 1.36 & 1.27 & 0.93 &  & 0.04 & 1.33 & 1.28 & 0.95 \\
  3 & $\taulin$ &  & 0.01 & 1.61 & 1.59 & 0.95 &  & -0.04 & 1.6 & 1.59 & 0.95 &  & -0.07 & 1.64 & 1.59 & 0.94 \\
  & $\ttaulin$ &  & -0.01 & 1.62 & 1.58 & 0.94 &  & -0.05 & 1.61 & 1.58 & 0.94 &  & -0.10 & 1.65 & 1.58 & 0.93 \\
  & $\tauker$ &  & 0.11 & 1.20 & 1.19 & 0.95 &  & 0.12 & 1.24 & 1.19 & 0.94 &  & 0.07 & 1.25 & 1.19 & 0.94 \\
  & $\ttauker$ &  & 0.18 & 1.23 & 1.18 & 0.94 &  & 0.19 & 1.27 & 1.18 & 0.93 &  & 0.14 & 1.27 & 1.18 & 0.93 \\
  & $\tauns$ &  & 0.04 & 1.62 & 1.52 & 0.93 &  & -0.02 & 1.62 & 1.52 & 0.93 &  & -0.06 & 1.66 & 1.52 & 0.92 \\
  & $\ttauns$ &  & 0.02 & 1.64 & 1.51 & 0.93 &  & -0.02 & 1.62 & 1.51 & 0.93 &  & -0.06 & 1.68 & 1.51 & 0.92 \\
  & $\taunnet$ &  & 0.01 & 1.39 & 1.38 & 0.95 &  & -0.01 & 1.43 & 1.38 & 0.94 &  & -0.06 & 1.43 & 1.38 & 0.94 \\
  & $\ttaunnet$ &  & -0.05 & 1.46 & 1.43 & 0.94 &  & -0.08 & 1.49 & 1.43 & 0.94 &  & -0.14 & 1.51 & 1.43 & 0.93 \\
  & $\taurf$ &  & 0.01 & 1.55 & 1.25 & 0.89 &  & -0.03 & 1.55 & 1.25 & 0.89 &  & -0.06 & 1.58 & 1.25 & 0.87 \\
  \vspace{0.25cm}
  & $\ttaurf$ &  & 0.01 & 1.68 & 1.31 & 0.88 &  & -0.05 & 1.67 & 1.31 & 0.88 &  & -0.07 & 1.71 & 1.31 & 0.86 \\
  4 & $\taulin$ &  & -0.04 & 3.78 & 3.75 & 0.95 &  & -0.11 & 3.69 & 3.75 & 0.95 &  & 0.13 & 3.90 & 3.75 & 0.94 \\
  & $\ttaulin$ &  & -0.06 & 3.59 & 3.57 & 0.95 &  & -0.11 & 3.54 & 3.57 & 0.95 &  & 0.12 & 3.71 & 3.58 & 0.94 \\
  & $\tauker$ &  & -0.01 & 3.75 & 3.65 & 0.94 &  & -0.10 & 3.67 & 3.65 & 0.95 &  & 0.17 & 3.86 & 3.66 & 0.94 \\
  & $\ttauker$ &  & -0.06 & 3.49 & 3.47 & 0.95 &  & -0.08 & 3.42 & 3.47 & 0.95 &  & 0.16 & 3.61 & 3.47 & 0.94 \\
  & $\tauns$ &  & -0.01 & 3.76 & 3.67 & 0.94 &  & -0.13 & 3.65 & 3.67 & 0.95 &  & 0.14 & 3.87 & 3.67 & 0.94 \\
  & $\ttauns$ &  & -0.07 & 3.49 & 3.47 & 0.95 &  & -0.09 & 3.43 & 3.47 & 0.95 &  & 0.15 & 3.61 & 3.47 & 0.94 \\
  & $\taunnet$ &  & -0.03 & 3.72 & 3.70 & 0.95 &  & -0.11 & 3.63 & 3.70 & 0.96 &  & 0.15 & 3.85 & 3.70 & 0.94 \\
  & $\ttaunnet$ &  & -0.07 & 3.49 & 3.47 & 0.95 &  & -0.10 & 3.43 & 3.47 & 0.94 &  & 0.15 & 3.61 & 3.48 & 0.94 \\
  & $\taurf$ &  & -0.04 & 3.79 & 3.11 & 0.89 &  & -0.11 & 3.70 & 3.11 & 0.90 &  & 0.13 & 3.90 & 3.12 & 0.88 \\
  & $\ttaurf$ &  & -0.08 & 3.51 & 3.44 & 0.94 &  & -0.10 & 3.43 & 3.44 & 0.94 &  & 0.15 & 3.63 & 3.45 & 0.94 \\
   \hline
		\end{tabular}}
\vskip 1.4mm
\begin{tablenotes}
\item Abbreviations: SD, standard deviation; SE, standard error; CP, coverage probability; Rand.: ran
    domization; nspline: natural spline; nnet: neural network; rf: random forest.
\end{tablenotes}
\end{threeparttable}
\end{table}

\subsection{High-dimensional data-generating models}
We consider high-dimensional data-generating models to determine whether the treatment effect estimators and variance estimators generated using different methods such as random forest and neural network are consistent in the high-dimensional case and to compare the efficiencies of the estimators. We generate additional covariates for Models 1--4. In total, we generate $p$ covariates.

Model 5:
Model 5 is based on Model 1, they have the same underlying model. The additional covariates are independent of $X_{ij}$, and follow a multivariate normal distribution with zero mean and a covariance matrix whose elements are all $0.2$ except for the diagonal elements, which have values of $1$.

Model 6:
Model 6 is based on Model 2, they have the same underlying model. The additional covariates are first generated as in Model 5, and we then randomly choose $\lfloor p/3 \rfloor$ covariates among the additional covariates and multiply them by $X_{i1}$ or $X_{i2}$ with equal probability to obtain the final high-dimensional covariates.

Model 7:
Model 7 is based on Model 3, they have the same underlying model. The additional covariates are independent of $X_{ij}$, and they follow a multivariate normal distribution with zero mean and the covariance matrix is a symmetric Toeplitz matrix whose first row is a geometric sequence with initial value $1$ and common ratio $0.5$.

Model 8:
Model 8 is based on Model 4, they have the same underlying model. The additional covariates are generated as in Model 6.

\subsection{High-dimensional simulation results}
Here, we present the simulation results of the high-dimensional methods with $p = 200$. The randomization settings are the same as those in the low-dimensional simulations. We apply the lasso, random forest, gradient boosting regression tree (gbrt), recursive partitioning and regression tree (rpart), and a neural network (nnet) with one hidden layer to all covariates to estimate $\hatha$.

\begin{table}
	\centering
	\caption{Simulated biases, standard deviations, standard errors, and coverage probabilities for different estimators and randomization methods under equal allocation ($\pi=1/2$) and high-dim-\\ensional data-generating models.}\label{tab::equalsimhigh1000}
	\vskip 1.4mm
	\begin{threeparttable}
		\setlength{\tabcolsep}{2pt}
\resizebox{0.8\textwidth}{90mm}{
		\begin{tabular}{llccccccccccccccc}
		\cline{1-17}
		&  & \makebox[0.025\textwidth][c]{} &
        \multicolumn{4}{c}{Complete Rand.} &
        \makebox[0.025\textwidth][c]{} &
        \multicolumn{4}{c}{Stratified Block Rand.} &
        \makebox[0.025\textwidth][c]{} &
        \multicolumn{4}{c}{Minimization} \\ \cline{3-17}
		Model &
		Estimator & &
		\multicolumn{1}{c}{Bias} &
		\multicolumn{1}{c}{SD} &
		\multicolumn{1}{c}{SE} &
        \multicolumn{1}{c}{CP} & &
		\multicolumn{1}{c}{Bias} &
		\multicolumn{1}{c}{SD} &
		\multicolumn{1}{c}{SE} &
        \multicolumn{1}{c}{CP} & &
		\multicolumn{1}{c}{Bias} &
		\multicolumn{1}{c}{SD} &
		\multicolumn{1}{c}{SE} &
        \multicolumn{1}{c}{CP}  \\
  \hline
  5 & $\taulasso$ &  &  0.07 & 2.95 & 2.92 & 0.95 &  & 0.03 & 2.96 & 2.91 & 0.95 &  & -0.01 & 2.98 & 2.91 & 0.94 \\
  & $\ttaulasso$ &  &  0.09 & 3.01 & 2.98 & 0.95 &  & 0.02 & 3.02 & 2.97 & 0.94 &  & -0.01 & 3.03 & 2.97 & 0.94 \\
  & $\taurfh$ &  &  -0.13 & 3.53 & 3.22 & 0.92 &  & -0.21 & 3.50 & 3.21 & 0.93 &  & -0.21 & 3.49 & 3.21 & 0.93 \\
  & $\ttaurfh$ &  &  -0.09 & 4.95 & 3.99 & 0.89 &  & -0.16 & 4.85 & 3.98 & 0.90 &  & -0.15 & 4.82 & 3.97 & 0.90 \\
  & $\taugbm$ &  &  0.08 & 3.06 & 3.01 & 0.95 &  & 0.04 & 3.06 & 3.01 & 0.94 &  & 0.01 & 3.04 & 3.01 & 0.95 \\
  & $\ttaugbm$ &  &  0.08 & 3.15 & 3.06 & 0.94 &  & 0.03 & 3.13 & 3.06 & 0.94 &  & 0.02 & 3.11 & 3.05 & 0.94 \\
  & $\taurpart$ &  &  -0.21 & 3.31 & 3.29 & 0.95 &  & -0.34 & 3.31 & 3.29 & 0.94 &  & -0.35 & 3.37 & 3.29 & 0.94 \\
  & $\ttaurpart$ &  &  0.88 & 4.30 & 4.21 & 0.94 &  & 0.69 & 4.27 & 4.17 & 0.94 &  & 0.78 & 4.29 & 4.17 & 0.94 \\
  & $\taunnet$ &  &  -1.33 & 6.97 & 5.48 & 0.87 &  & -1.54 & 7.06 & 5.48 & 0.86 &  & -1.47 & 7.08 & 5.50 & 0.87 \\
  \vspace{0.25cm}
  & $\ttaunnet$ &  &  -0.40 & 8.70 & 6.72 & 0.86 &  & -0.42 & 8.68 & 6.71 & 0.87 &  & -0.60 & 8.58 & 6.70 & 0.87 \\
  6 & $\taulasso$ &  &  -0.09 & 1.59 & 1.55 & 0.94 &  & -0.09 & 1.56 & 1.55 & 0.95 &  & -0.06 & 1.54 & 1.55 & 0.95 \\
  & $\ttaulasso$ &  &  -0.42 & 1.68 & 1.56 & 0.92 &  & -0.41 & 1.64 & 1.55 & 0.93 &  & -0.39 & 1.63 & 1.56 & 0.93 \\
  & $\taurfh$ &  &  -0.22 & 1.41 & 1.28 & 0.92 &  & -0.22 & 1.35 & 1.28 & 0.93 &  & -0.20 & 1.37 & 1.28 & 0.93 \\
  & $\ttaurfh$ &  &  -0.48 & 1.55 & 1.29 & 0.88 &  & -0.48 & 1.47 & 1.29 & 0.90 &  & -0.44 & 1.49 & 1.29 & 0.89 \\
  & $\taugbm$ &  &  -0.04 & 1.32 & 1.29 & 0.94 &  & -0.03 & 1.27 & 1.29 & 0.96 &  & -0.02 & 1.30 & 1.29 & 0.95 \\
  & $\ttaugbm$ &  &  -0.18 & 1.36 & 1.29 & 0.93 &  & -0.19 & 1.31 & 1.29 & 0.95 &  & -0.17 & 1.32 & 1.29 & 0.94 \\
  & $\taurpart$ &  &  -0.11 & 1.38 & 1.36 & 0.95 &  & -0.11 & 1.34 & 1.35 & 0.96 &  & -0.09 & 1.35 & 1.35 & 0.95 \\
  & $\ttaurpart$ &  &  -0.01 & 1.50 & 1.46 & 0.94 &  & -0.02 & 1.42 & 1.45 & 0.96 &  & 0.01 & 1.46 & 1.46 & 0.95 \\
  & $\taunnet$ &  &  0.08 & 2.49 & 1.92 & 0.87 &  & 0.09 & 2.46 & 1.92 & 0.87 &  & 0.24 & 2.46 & 1.92 & 0.87 \\
  \vspace{0.25cm}
  & $\ttaunnet$ &  &  -0.73 & 2.53 & 1.97 & 0.86 &  & -0.73 & 2.55 & 1.96 & 0.85 &  & -0.70 & 2.62 & 1.96 & 0.85 \\
  7 & $\taulasso$ &  &  0.00 & 1.67 & 1.68 & 0.95 &  & 0.00 & 1.69 & 1.68 & 0.95 &  & -0.09 & 1.62 & 1.68 & 0.96 \\
  & $\ttaulasso$ &  &  -0.02 & 1.70 & 1.71 & 0.95 &  & -0.01 & 1.73 & 1.70 & 0.95 &  & -0.10 & 1.64 & 1.71 & 0.96 \\
  & $\taurfh$ &  &  0.26 & 1.67 & 1.22 & 0.84 &  & 0.28 & 1.70 & 1.22 & 0.84 &  & 0.20 & 1.61 & 1.22 & 0.86 \\
  & $\ttaurfh$ &  &  0.31 & 1.77 & 1.27 & 0.83 &  & 0.32 & 1.80 & 1.27 & 0.82 &  & 0.24 & 1.70 & 1.27 & 0.85 \\
  & $\taugbm$ &  &  -0.01 & 1.64 & 1.50 & 0.92 &  & 0.01 & 1.68 & 1.49 & 0.92 &  & -0.08 & 1.59 & 1.50 & 0.93 \\
  & $\ttaugbm$ &  &  -0.02 & 1.66 & 1.42 & 0.91 &  & -0.01 & 1.70 & 1.41 & 0.90 &  & -0.09 & 1.63 & 1.42 & 0.91 \\
  & $\taurpart$ &  &  -0.05 & 1.59 & 1.50 & 0.93 &  & -0.01 & 1.61 & 1.50 & 0.94 &  & -0.07 & 1.58 & 1.50 & 0.94 \\
  & $\ttaurpart$ &  &  -0.02 & 1.64 & 1.52 & 0.93 &  & -0.04 & 1.67 & 1.52 & 0.93 &  & -0.10 & 1.65 & 1.52 & 0.93 \\
  & $\taunnet$ &  &  0.47 & 2.62 & 1.88 & 0.84 &  & 0.44 & 2.68 & 1.88 & 0.84 &  & 0.31 & 2.67 & 1.88 & 0.82 \\
  \vspace{0.25cm}
  & $\ttaunnet$ &  &  0.88 & 2.77 & 2.00 & 0.81 &  & 0.94 & 2.79 & 2.00 & 0.82 &  & 0.90 & 2.67 & 2.00 & 0.83 \\
  8 & $\taulasso$ &  &  -0.07 & 3.76 & 3.77 & 0.94 &  & -0.03 & 3.74 & 3.77 & 0.95 &  & 0.04 & 3.85 & 3.77 & 0.94 \\
  & $\ttaulasso$ &  &  -0.13 & 3.65 & 3.69 & 0.94 &  & -0.05 & 3.67 & 3.70 & 0.95 &  & 0.04 & 3.79 & 3.70 & 0.94 \\
  & $\taurfh$ &  &  0.49 & 3.62 & 2.83 & 0.87 &  & 0.50 & 3.55 & 2.83 & 0.88 &  & 0.55 & 3.68 & 2.84 & 0.86 \\
  & $\ttaurfh$ &  &  0.59 & 3.54 & 3.36 & 0.93 &  & 0.65 & 3.50 & 3.37 & 0.94 &  & 0.71 & 3.63 & 3.37 & 0.92 \\
  & $\taugbm$ &  &  0.04 & 3.67 & 3.36 & 0.92 &  & 0.05 & 3.58 & 3.36 & 0.93 &  & 0.07 & 3.71 & 3.36 & 0.92 \\
  & $\ttaugbm$ &  &  -0.09 & 3.45 & 3.47 & 0.95 &  & 0.01 & 3.39 & 3.47 & 0.95 &  & 0.09 & 3.55 & 3.47 & 0.94 \\
  & $\taurpart$ &  &  0.03 & 3.81 & 3.44 & 0.92 &  & 0.02 & 3.75 & 3.45 & 0.92 &  & 0.07 & 3.91 & 3.45 & 0.91 \\
  & $\ttaurpart$ &  &  -0.08 & 3.49 & 3.51 & 0.94 &  & 0.01 & 3.43 & 3.51 & 0.95 &  & 0.09 & 3.58 & 3.51 & 0.94 \\
  & $\taunnet$ &  &  0.09 & 4.84 & 3.79 & 0.87 &  & -0.05 & 4.76 & 3.81 & 0.89 &  & 0.19 & 4.84 & 3.80 & 0.87 \\
  & $\ttaunnet$ &  &  1.25 & 4.96 & 3.79 & 0.86 &  & 1.25 & 4.80 & 3.79 & 0.87 &  & 1.21 & 4.98 & 3.79 & 0.86 \\
   \hline
		\end{tabular}}
\vskip 1.4mm
\begin{tablenotes}
\item Abbreviations: SD, standard deviation; SE, standard error; CP, coverage probability; Rand.: randomization; rf: random forest; gbrt: gradient boosting regression tree; rpart: recursive partitioning and regression tree; nnet: neural network.
\end{tablenotes}
\end{threeparttable}
\end{table}

From Table~\ref{tab::equalsimhigh1000}, we see that in the high-dimensional cases, the treatment effect estimators still behave similarly under different randomization methods. Under the considered scenarios, they all have small biases, except for $\taunn$. When the data-generating model is a high-dimensional linear model (Model 5), $\taulasso$ has standard deviations similar to those of $\taulin$ under the corresponding low-dimensional model (Model 1), indicating that lasso makes good prediction and $\taulasso$ achieves the optimal efficiency. All other treatment effect estimators have larger standard deviations than $\taulasso$. When the data-generating model is nonlinear (Models 6, 7, and 8), the tree-based methods have good fitting results. The performances of the treatment effect estimators vary from model to model. For example, among stratum-common estimators, $\taugbm$ has the smallest standard deviations in Model 6, $\taurpart$ has the smallest standard deviations in Model 7, and $\taurfh$ has the smallest standard deviations in Model 8. In contrast, $\taunn$ always has relatively large standard deviations, suggesting that, unlike the low-dimensional cases, the neural network does not estimate the true model as well as the tree-based methods. Additionally, similar to the low-dimensional cases, the stratum-specific estimators have smaller standard deviations than the stratum-common estimators only when the data-generating model is stratum-specific (Model 8).

From the perspective of inference, no machine learning-adjusted estimator always has a valid 95\% coverage probability. However, for nonlinear data-generating models (Models 6, 7, and 8), the machine learning-adjusted estimators have smaller standard deviations than the lasso-adjusted estimators. This implies that the machine learning methods converge to certain functions, but because we reuse samples in the estimation process, complex dependencies are introduced. Thus, the convergence properties required by Assumption 4 may not be satisfied. This motivates us to use the sample splitting technique to alleviate the requirements, as discussed in Section~\ref{subsec::sample}.

\begin{table}
	\centering
	\caption{Simulated biases, standard deviations, standard errors, and coverage probabilities for different sample splitting estimators and randomization methods under equal allocation ($\pi=1/2$) and high-dimensional data-generating models.}\label{tab::equalsimhighds1000}
	\vskip 1.4mm
	\begin{threeparttable}
		\setlength{\tabcolsep}{2pt}
\resizebox{0.8\textwidth}{90mm}{
		\begin{tabular}{llccccccccccccccc}
		\cline{1-17}
		&  & \makebox[0.025\textwidth][c]{} &
        \multicolumn{4}{c}{Complete Rand.} &
        \makebox[0.025\textwidth][c]{} &
        \multicolumn{4}{c}{Stratified Block Rand.} &
        \makebox[0.025\textwidth][c]{} &
        \multicolumn{4}{c}{Minimization} \\ \cline{3-17}
		Model &
		Estimator & &
		\multicolumn{1}{c}{Bias} &
		\multicolumn{1}{c}{SD} &
		\multicolumn{1}{c}{SE} &
        \multicolumn{1}{c}{CP} & &
		\multicolumn{1}{c}{Bias} &
		\multicolumn{1}{c}{SD} &
		\multicolumn{1}{c}{SE} &
        \multicolumn{1}{c}{CP} & &
		\multicolumn{1}{c}{Bias} &
		\multicolumn{1}{c}{SD} &
		\multicolumn{1}{c}{SE} &
        \multicolumn{1}{c}{CP}  \\
  \hline
  5 & $\taulassoss$ &  &  0.04 & 2.97 & 2.89 & 0.95 &  & 0.04 & 2.98 & 2.89 & 0.94 &  & 0.00 & 3.01 & 2.89 & 0.94 \\
  & $\ttaulassoss$ &  &  0.07 & 3.08 & 3.00 & 0.95 &  & 0.04 & 3.09 & 3.00 & 0.94 &  & 0.01 & 3.10 & 3.00 & 0.94 \\
  & $\taurfhss$ &  &  0.10 & 3.81 & 3.67 & 0.94 &  & 0.03 & 3.77 & 3.66 & 0.94 &  & 0.02 & 3.74 & 3.66 & 0.94 \\
  & $\ttaurfhss$ &  &  0.10 & 5.45 & 5.27 & 0.94 &  & 0.06 & 5.39 & 5.25 & 0.94 &  & 0.04 & 5.33 & 5.25 & 0.95 \\
  & $\taugbmss$ &  &  0.08 & 3.02 & 2.94 & 0.94 &  & 0.03 & 3.06 & 2.94 & 0.94 &  & 0.01 & 3.03 & 2.94 & 0.94 \\
  & $\ttaugbmss$ &  &  0.12 & 3.67 & 3.54 & 0.94 &  & 0.02 & 3.64 & 3.52 & 0.94 &  & 0.07 & 3.59 & 3.52 & 0.94 \\
  & $\taurpartss$ &  &  0.12 & 3.40 & 3.33 & 0.94 &  & 0.07 & 3.41 & 3.32 & 0.94 &  & 0.00 & 3.42 & 3.32 & 0.94 \\
  & $\ttaurpartss$ &  &  0.17 & 4.88 & 4.75 & 0.94 &  & 0.01 & 4.88 & 4.73 & 0.94 &  & 0.10 & 4.80 & 4.72 & 0.94 \\
  & $\taunnetss$ &  &  0.14 & 6.97 & 6.72 & 0.94 &  & 0.34 & 6.81 & 6.71 & 0.95 &  & -0.06 & 6.86 & 6.72 & 0.94 \\
  \vspace{0.25cm}
  & $\ttaunnetss$ &  &  0.14 & 8.62 & 8.37 & 0.95 &  & 0.24 & 8.67 & 8.35 & 0.94 &  & 0.12 & 8.65 & 8.36 & 0.94 \\
  6 & $\taulassoss$ &  & -0.03 & 1.61 & 1.55 & 0.94 &  & -0.02 & 1.57 & 1.55 & 0.95 &  & 0.01 & 1.57 & 1.55 & 0.95 \\
  & $\ttaulassoss$ &  &  -0.03 & 1.80 & 1.74 & 0.94 &  & -0.01 & 1.72 & 1.73 & 0.95 &  & 0.02 & 1.73 & 1.73 & 0.95 \\
  & $\taurfhss$ &  &  -0.02 & 1.44 & 1.38 & 0.94 &  & -0.02 & 1.37 & 1.38 & 0.95 &  & 0.01 & 1.39 & 1.38 & 0.95 \\
  & $\ttaurfhss$ &  &  -0.01 & 1.58 & 1.52 & 0.94 &  & -0.01 & 1.51 & 1.52 & 0.95 &  & 0.01 & 1.53 & 1.52 & 0.95 \\
  & $\taugbmss$ &  &  -0.02 & 1.34 & 1.29 & 0.94 &  & -0.02 & 1.28 & 1.29 & 0.95 &  & 0.00 & 1.31 & 1.29 & 0.95 \\
  & $\ttaugbmss$ &  &  -0.02 & 1.47 & 1.42 & 0.94 &  & -0.02 & 1.40 & 1.41 & 0.95 &  & 0.01 & 1.42 & 1.41 & 0.95 \\
  & $\taurpartss$ &  &  -0.02 & 1.40 & 1.35 & 0.94 &  & -0.01 & 1.35 & 1.35 & 0.95 &  & 0.00 & 1.38 & 1.35 & 0.94 \\
  & $\ttaurpartss$ &  &  -0.01 & 1.55 & 1.50 & 0.94 &  & -0.02 & 1.48 & 1.50 & 0.96 &  & 0.03 & 1.53 & 1.50 & 0.94 \\
  & $\taunnetss$ &  &  -0.03 & 2.32 & 2.28 & 0.95 &  & -0.05 & 2.36 & 2.28 & 0.94 &  & 0.04 & 2.35 & 2.28 & 0.95 \\
  \vspace{0.25cm}
  & $\ttaunnetss$ &  &  -0.06 & 2.40 & 2.37 & 0.95 &  & -0.04 & 2.41 & 2.36 & 0.95 &  & 0.01 & 2.38 & 2.36 & 0.94 \\
  7 & $\taulassoss$ &  &  0.01 & 1.68 & 1.68 & 0.95 &  & 0.02 & 1.71 & 1.68 & 0.95 &  & -0.08 & 1.63 & 1.68 & 0.95 \\
  & $\ttaulassoss$ &  &  0.01 & 1.72 & 1.72 & 0.95 &  & 0.02 & 1.75 & 1.71 & 0.95 &  & -0.08 & 1.66 & 1.72 & 0.95 \\
  & $\taurfhss$ &  &  0.01 & 1.68 & 1.67 & 0.95 &  & 0.01 & 1.70 & 1.66 & 0.95 &  & -0.06 & 1.63 & 1.67 & 0.95 \\
  & $\ttaurfhss$ &  &  0.02 & 1.78 & 1.77 & 0.95 &  & 0.02 & 1.81 & 1.77 & 0.95 &  & -0.07 & 1.73 & 1.77 & 0.95 \\
  & $\taugbmss$ &  &  -0.01 & 1.57 & 1.57 & 0.95 &  & 0.01 & 1.61 & 1.57 & 0.95 &  & -0.07 & 1.55 & 1.57 & 0.95 \\
  & $\ttaugbmss$ &  &  0.00 & 1.63 & 1.64 & 0.95 &  & 0.01 & 1.67 & 1.64 & 0.95 &  & -0.07 & 1.60 & 1.64 & 0.96 \\
  & $\taurpartss$ &  &  -0.02 & 1.64 & 1.62 & 0.95 &  & 0.03 & 1.64 & 1.62 & 0.95 &  & -0.06 & 1.61 & 1.62 & 0.95 \\
  & $\ttaurpartss$ &  &  0.01 & 1.68 & 1.67 & 0.95 &  & -0.02 & 1.73 & 1.67 & 0.94 &  & -0.07 & 1.65 & 1.68 & 0.95 \\
  & $\taunnetss$ &  &  -0.04 & 2.33 & 2.34 & 0.95 &  & -0.01 & 2.40 & 2.34 & 0.95 &  & -0.09 & 2.26 & 2.34 & 0.95 \\
  \vspace{0.25cm}
  & $\ttaunnetss$ &  &  0.02 & 2.52 & 2.47 & 0.95 &  & 0.05 & 2.57 & 2.47 & 0.93 &  & -0.14 & 2.44 & 2.47 & 0.95 \\
  8 & $\taulassoss$ &  &  -0.03 & 3.79 & 3.77 & 0.94 &  & -0.01 & 3.76 & 3.77 & 0.95 &  & 0.06 & 3.88 & 3.77 & 0.94 \\
  & $\ttaulassoss$ &  &  -0.07 & 3.67 & 3.69 & 0.94 &  & 0.01 & 3.68 & 3.70 & 0.95 &  & 0.09 & 3.81 & 3.70 & 0.94 \\
  & $\taurfhss$ &  &  -0.02 & 3.67 & 3.67 & 0.95 &  & -0.02 & 3.61 & 3.67 & 0.95 &  & 0.05 & 3.75 & 3.67 & 0.95 \\
  & $\ttaurfhss$ &  &  -0.06 & 3.54 & 3.54 & 0.94 &  & 0.01 & 3.49 & 3.54 & 0.95 &  & 0.06 & 3.63 & 3.55 & 0.94 \\
  & $\taugbmss$ &  &  -0.01 & 3.72 & 3.71 & 0.94 &  & 0.02 & 3.65 & 3.71 & 0.95 &  & 0.02 & 3.80 & 3.71 & 0.94 \\
  & $\ttaugbmss$ &  &  -0.08 & 3.47 & 3.47 & 0.94 &  & 0.01 & 3.42 & 3.47 & 0.95 &  & 0.09 & 3.56 & 3.48 & 0.94 \\
  & $\taurpartss$ &  &  0.02 & 3.82 & 3.83 & 0.95 &  & 0.03 & 3.77 & 3.83 & 0.95 &  & 0.02 & 3.95 & 3.83 & 0.94 \\
  & $\ttaurpartss$ &  &  -0.06 & 3.52 & 3.53 & 0.94 &  & 0.02 & 3.48 & 3.53 & 0.95 &  & 0.09 & 3.64 & 3.53 & 0.94 \\
  & $\taunnetss$ &  &  -0.03 & 4.50 & 4.49 & 0.94 &  & -0.02 & 4.44 & 4.49 & 0.95 &  & 0.03 & 4.70 & 4.50 & 0.93 \\
  & $\ttaunnetss$ &  &  -0.03 & 4.26 & 4.21 & 0.94 &  & -0.04 & 4.18 & 4.21 & 0.95 &  & 0.12 & 4.39 & 4.22 & 0.94 \\
   \hline
		\end{tabular}}
\vskip 1.4mm
\begin{tablenotes}
\item Abbreviations: SD, standard deviation; SE, standard error; CP, coverage probability; Rand.: randomization; rf: random forest; gbrt: gradient boosting regression tree; rpart: recursive partitioning and regression tree; nnet: neural network; ss: sample splitting.
\end{tablenotes}
\end{threeparttable}
\end{table}

Table~\ref{tab::equalsimhighds1000} presents the simulation results obtained using the sample splitting and cross-fitting techniques (see Algorithm~\ref{alg1}). First, all treatment effect estimators tend to have smaller biases under considered scenarios with the used techniques. Particularly, in Model 8, under simple randomization, the absolute value of the bias of $\ttaunnet$ is 0.03, which is $2.4\%$ of the bias of the corresponding $\ttaunn$ under the low-dimensional settings. In addition, the treatment estimators have standard deviations similar to those obtained without sample splitting, indicating the regain of efficiency with cross-fitting. There is still no one covariate adjustment method that always yields the best treatment effect estimator. However, $\taunnetss$ and $\ttaunnetss$ continues to have the largest standard deviations.

In terms of statistical inference, we see that all treatment effect estimators have coverage probabilities of approximately $95\%$. In other words, sample splitting can effectively alleviate the requirement on covariate adjustment methods (Assumption~\ref{assum4}) and thus allow valid inference.

\section{Discussion and Practical Recommendations}\label{sec::discussion}
In this paper, we inherited the framework of \cite{Liu2020Lasso} and proved the asymptotic property of the oracle treatment effect estimator when the forms of the covariate adjustment functions are known. We reviewed the assumptions presented in \cite{Liu2020Lasso} that adjusting methods need to satisfy to realize valid inference once the covariate adjustment functions are plugged in. We presented a detailed verification of the assumptions' satisfiability for local linear kernel. Following the asymptotic distribution, valid confidence intervals and tests could be constructed for the average treatment effect. In the case of high-dimensional covariates, we proposed estimators using machine learning methods and the sample splitting technique, which had efficient and robust practical performances.

According to the theoretical verifications and simulation results and considering simplicity, robustness, and efficiency, our recommendations to practitioners are as follows. If a strong linear relationship between covariates and outcomes is observed, we recommend linear regression or lasso for covariate adjustment, because these methods can readily and quickly lead to a valid and efficient treatment effect estimator. We consider using other methods when the relationship between covariates and outcomes is complex. In low-dimensional cases, we can use the local linear kernel or smoothing spline to adjust covariates for valid inference. In high-dimensional cases, we can choose appropriate machine learning methods by considering the background information and distributions of covariates and outcomes. Moreover, sample splitting and cross-fitting techniques should be used to obtain valid inferences and regain efficiency.

Independently of our work but at the same time, \cite{Rafi2023} and \cite{Bannick2023} also studied how to improve the efficiency of the treatment effect estimator under stratified randomization and used sample splitting and cross-fitting to alleviate the requirements on the estimation methods. While we shared similar ideas, our considered estimators, adjusting methods and contributions are different. \cite{Rafi2023} explored the semiparametric efficiency bound and proposed the second moment convergence assumption on the estimation function under sample splitting for the consistency and efficiency of the treatment effect estimator. Inspired by his proofs, we provided the theoretical justification of our proposed sample splitting estimator based on the Assumption~\ref{assum::ss}. It should be noted that our sample splitting algorithm is slightly different from that used by \cite{Rafi2023}. Our algorithm is carried out on the whole sample, while \cite{Rafi2023} performed the sample splitting process within each stratum. Importantly, although our theoretical results are inspired by \cite{Rafi2023}, we have uniquely introduced the sample splitting algorithm and estimator in our research. Moreover, \cite{Rafi2023} proved that when using sample splitting and cross-fitting, the estimator adjusted by Nadaraya-Watson kernel regression can achieve the efficiency bound. In this paper, we proved that the treatment effect estimator adjusted by local linear kernel regression can also attain the corresponding asymptotic property without sample splitting. By comparing the asymptotic variances, we can conclude that our proposed estimator adjusted by local linear kernel can also achieve the efficiency bound while retaining high computational efficiency. In addition, we provide consistent variance estimator for the average treatment effect, hence solving the problem of constructing valid inference procedures which is left in \cite{Rafi2023}. \cite{Bannick2023} considered the Donsker condition for parametric and nonparametric methods in low-dimensional cases, which is an extension of \cite{Guo2023} and \cite{Zheng2019}, while we directly verified the asymptotic properties for kernel regression. In high-dimensional cases, \cite{Bannick2023} also imposed an $L_2$ condition and established the theoretical properties of the cross-fitted estimator, although the exact form of the condition and the estimator are slightly different than ours. In numerical experiments, \cite{Bannick2023} applied their results to the generalized linear model and random forest. Meanwhile, we considered a broad class of adjusting methods and they all obtain good empirical performances.

\bibliographystyle{apalike}
\bibliography{refs}


\appendix



\section{Useful Lemmas}
We first introduce the following lemmas that are useful for our proofs.

\begin{lemma}
\label{lem1}
Let $V_i = f(Y_i(1),Y_i(0),B_i, \bx_i)$ for some measurable function   $f(\cdot)$  such that  $E( |V_i| ) < \infty$. Under Assumptions~\ref{assum1}--\ref{assum3},
$$
\frac{1}{n} \sumi A_i V_i  \xrightarrow{P} \pi  E(V_1).
$$
\end{lemma}

\begin{lemma}
\label{lem2}
Under Assumptions \ref{assum1}--\ref{assum3}, we have
$$
\frac{\nt}{n} \xrightarrow{P} \pi, \quad \pink = \frac{\nkt}{\nk} \xrightarrow{P} \pi, \quad \frac{\nkt}{n}  \xrightarrow{P} \pi \pk, \quad \pnk = \frac{\nk}{n}  \xrightarrow{P} \pk,
$$
$$
\frac{\nc}{n} \xrightarrow{P} 1 - \pi, \quad \frac{\nkc}{\nk} \xrightarrow{P} 1 - \pi,  \quad \frac{\nkc}{n}  \xrightarrow{P} ( 1 -  \pi ) \pk.
$$
\end{lemma}

\begin{lemma}
\label{lem3}
Let $V_i = f(Y_i(1),Y_i(0),B_i, \bx_i)$ for some measurable function $f(\cdot)$  such that  $E ( V_i^2 ) < \infty$. Under Assumptions~\ref{assum1}--\ref{assum3},
$$
\sumk \pnk \cdot \frac{1}{\nkt} \sumik  A_i ( V_i - \bar{V}_{[k]1} )^2  \xrightarrow{P}  \sigma^2_{V_i - E(V_i \mid B_i ) },
$$
$$
\sumk \pnk \cdot \frac{1}{\nkc} \sumik  (1 - A_i )  ( V_i - \bar{V}_{[k]0} )^2  \xrightarrow{P}  \sigma^2_{V_i - E(V_i \mid B_i ) }.
$$
\end{lemma}

The above lemmas are the same as those in \cite{Liu2020Lasso}, and Lemma~\ref{lem1} is a generalized version of what has been proved in \cite{Bugni2019} for $V_i = f(Y_i(1),Y_i(0),B_i)$ (see Lemma C.4). Lemma~\ref{lem2} can be obtained directly from  the weak law of large numbers and the above Lemma~\ref{lem1}. Lemma \ref{lem3} can be obtained from the proof of Lemma 7 in \cite{Ma2020Regression}. We omit the proofs of these lemmas.

\section{Proof of Proposition~\ref{Prop1}}

\begin{proof}
Let $
\hk = (1-\pi) \hts + \pi \hcs, \
\barhkt = \nkt^{-1} \sum_{i \in [k]} A_i \hk$
and $ \barhkc = \nkc^{-1} \sum_{i \in [k]} (1 - A_i) \hk$. Then
\begin{eqnarray}\label{eq3}
\nonumber  \tauor &=& \sumk \pnk \bigg[\Big\{\YkThat  - \sumik \frac{A_i - \pink}{\nk \pink}\hts\Big\} \\
\nonumber & & - \Big\{\YkChat  + \sumik \frac{A_i - \pink}{\nk(1-\pink)}\hcs\Big\}\bigg] \\
\nonumber &=& \sumk \frac{\pnk}{\nk} \sumik \Big\{\frac{A_i Y_i}{\pink} - \frac{A_i - \pink}{\pink} \hts \\
\nonumber & & - \frac{(1 - A_i)Y_i}{1 - \pink} - \frac{A_i - \pink}{1 - \pink}\hcs \Big\} \\
\nonumber   &=& \sumk \frac{\pnk}{\nk} \sumik \Big[ \frac{A_i}{\pink} \big\{Y_i - \hk \big\} -  \frac{1 - A_i}{1 - \pink} \big\{Y_i - \hk \big\}\Big] \\
\nonumber && + \sumk \frac{\pnk}{\nk} \sumik \Big[\frac{A_i}{\pink}(\pink - \pi) \big\{\hts - \hcs \big\} \\
\nonumber &&  - \frac{1-A_i}{1-\pink}(\pink - \pi) \big\{\hts - \hcs \big\} \Big] \\
\nonumber   &=& \sumk \pnk \Big[\YkThat  - \barhkt - \big\{\YkChat  - \barhkc \big\} \Big] \\
\nonumber && + \sumk \pnk (\pink - \pi) \frac{1}{\nkt} \sumik A_i \big\{\hts - \hcs \big\} \\
 &&  - \sumk \pnk (\pink - \pi) \frac{1}{\nkc} \sumik (1-A_i) \big\{\hts - \hcs \big\}.
\end{eqnarray}
The equality in equation~\eqref{eq3} is because
\begin{center}
$\begin{aligned}
\YkThat  &= \frac{1}{\nkt} \sumik A_i Y_i = \frac{1}{\nk \pink} \sumik A_i Y_i, \\
\YkChat  &= \frac{1}{\nkc} \sumik (1-A_i) Y_i = \frac{1}{\nk (1-\pink)} \sumik (1-A_i) Y_i.
\end{aligned}$
\end{center}

For $i \in [k]$, denote the transformed outcome as
$$r_i(a) = Y_i (a) - [(1-\pi) \hts + \pi \hcs],$$
then $\tauor$ is the stratified difference-in-means estimator applied to the transformed outcomes $r_i(a), \ a = 0,1$, which satisfy
$$
E\{r_i(1) - r_i(0)\} = \sumk p_{[k]} E\{Y_i(1) - Y_i(0)|B_i = k\} = E\{Y_i(1) - Y_i(0)\} = \tau.$$

Since $E\{Y^2_i(a)\} < \infty$ and $E\{ h^2_{[k]}(X_i,a) \} < \infty$, then $E\{ r^2_i(a) \} < \infty$. As a result, according to Proposition 1 in \cite{Liu2020Lasso}, $\sumk \pnk \big[\YkThat  - \barhkt - \{\YkChat  - \barhkc\} \big]$ is asymptotically normal with mean $\tau$ and variance $\zeta^2_{r}(\pi) + \zeta^2_{Hr}$. Then, it suffices to show that the last two terms in equation~(\ref{eq3}) are negligible.

Under the second moment conditions on $h_{[k]}(X_i, a)$, applying Proposition 1 in \cite{Liu2020Lasso} to each stratum with outcomes $\hts - \hcs$, we have $(1/\nkt) \sumik A_i \{\hts - \hcs\} - (1/\nkc) \sumik (1-A_i) \{\hts - \hcs\} = O_P(n^{-1/2})$. Together with $\pink - \pi = o_P(1)$, we have the desired term is $o_P(n^{-1/2})$.
\end{proof}

\section{Proof of Local Linear Kernel}

We denote $K_H(u) = |H|^{-1/2}K(H^{-1/2}u)$, where $K(\cdot)$ is the symmetric density kernel function used in local linear kernel, $H$ is a $d \times d$ symmetric positive definite matrix depending on $n$. $H^{1/2}$ is called the bandwidth matrix. Let $D_g(x)$ denote the $d\times 1$ vector of first-order partial derivatives and $\mathcal{H}_g(x)$ denote the $d \times d$ Hessian matrix of a sufficiently smooth $d$-variate function $g$ at $x$. Let $\textbf{1}$ denote a generic matrix having each entry equal to $1$. If $U_n$ is a random matrix, then $O_P(U_n)$ and $o_P(U_n)$ are to be taken componentwise.

\begin{lemma}
\label{lem::ks}
  Suppose that

  (i) $X_1, \dots, X_n$ are i.i.d. (d-)dimensional variables with continuous probability density function $f(\cdot)$, and $f(\cdot)$ has a compact support set on $R^d$;

  (ii) $Y_i = m(X_i) + \nu^{1/2}(X_i)\varepsilon_i, \ i= 1, \dots, n$, where $\nu(x) = \Var(Y|X = x)>0$ is continuous,  $\varepsilon_i$'s are mutually independent random variables with $E(\varepsilon_i) = 0, \ \textnormal{Var}(\varepsilon_i) = 1$, and $\varepsilon_i$'s are independent of $X_i$'s. All second-order derivatives of $m(\cdot)$ are continuous;

  (iii) The kernel K is a compactly supported, bounded kernel such that $\int uu^\T K(u) du = \mu_2(K)I$, where $\mu_2(K) \neq 0$ is a scalar and $I$ is the $d \times d$ identity matrix. In addition, all odd-order moments of $K$ vanish, that is, $\int u_1^{l1}\cdots u_d^{l_d} K(u) du = 0$ for all nonnegative integers $l_1,\dots,\l_d$ such that their sum is odd;

  (iv) $n^{-1}|H|^{-1}$ and each entry of $H$ tend to zero as $n \to \infty$, with $H$ remaining symmetric and positive definite. Moreover, there is a fixed constant $L$ such that the condition number of $H$ is at most $L$ for all $n$.

  Denote $\mhxq$ as the local linear smoother of $m(\cdot)$ on point $x$ based on $X_1, \dots, X_n$, using symmetric density kernel and bandwidth matrix $H^{1/2}$. Then
  $$E\left[\left\{\mhxq - m(x)\right\}^2|X_1,\dots,X_n\right] \xrightarrow{P} 0.$$
\end{lemma}

\begin{proof}
  Denote $$M_0 = (m(X_1), \dots, m(X_n))^{\T}, \quad Y = (Y_1, \dots, Y_n)^{\T},$$
  $$W_x = \textnormal{diag}(K_H(x - X_1), \dots, K_H(x - X_n)),$$
$$N_x = \begin{pmatrix}
        1 & (X_1 - x)^{\T} \\
        \vdots & \vdots \\
        1 & (X_n - x)^{\T}
      \end{pmatrix}, \quad
V = \begin{pmatrix}
        \nu(X_1) &  &  \\
         & \ddots &  \\
         &  & \nu(X_n)
      \end{pmatrix}.$$
By the definition of the local linear kernel smoother, for given $x$, we have

$$ \mhxq = e_1^T(N_x ^\T W_x N_x )^{-1}N_x^TW_x Y,$$
where $e_1$ is a $d$-dimensional vector with first element being $1$ and the remaining elements being $0$. Taking expectation with respect to $Y$, we have
\begin{equation}\label{eqn::C1}
E\left\{\mhxq \mid X_{1}, \ldots, X_{n}\right\}=e_{1}^\T\left(N_x ^\T W_x N_x \right)^{-1} N_x ^\T W_x  M_0.
\end{equation}

Let $Q_m(x)$ be the $n \times 1$ vector given by
\begin{equation}\label{eqn::C2}
Q_{m}(x)=\left[\left(X_{1}-x \right)^\T \mathcal{H}_{m}(x)\left(X_{1}-x\right), \ldots,\left(X_{n}-x\right)^\T \mathcal{H}_{m}(x)\left(X_{n}-x\right)\right]^\T.
\end{equation}
Then Taylor's expansion implies that
\begin{equation}\label{eqn::C3}
  M_0=N_x \left(\begin{array}{c}
\mxq \\
D_{m}(x)
\end{array}\right)+\frac{1}{2} Q_{m}(x)+R_{m}(x),
\end{equation}
where $R_m(x)$ is a vector of Taylor series remainder terms. As \cite{Ruppert1994} stated, when $R_m(x)$ is pre-multiplied by $e_1^\T (N_x^\T W_x N_x)^{-1}N_x^\T W_x$, the resulting scalar is $o_P\{\textnormal{tr}(H)\}$. Then by equations~(\ref{eqn::C1})--(\ref{eqn::C3}),
\begin{eqnarray*}
&&E\left\{\mhxq -\mxq \mid X_{1}, \ldots, X_{n}\right\} \\
&=&e_{1}^\T\left(N_x^\T W_x  N_x\right)^{-1} N_x^\T W_x  \left\{N_x\left(\begin{array}{c}
\mxq \\
D_{m}(x)
\end{array}\right)+\frac{1}{2} Q_{m}(x)+R_{m}(x) \right\} - \mxq \\
&=&e_{1}^\T\left(N_x^\T W_x  N_x\right)^{-1} N_x^\T W_x N_x \left(\begin{array}{c}
\mxq \\
D_{m}(x)
\end{array}\right) \\
&& + e_{1}^\T\left(N_x^\T W_x  N_x\right)^{-1} N_x^\T W_x  \left\{\frac{1}{2} Q_{m}(x)+R_{m}(x) \right\} - \mxq \\
&=& e_{1}^\T\left(\begin{array}{c}
\mxq \\
D_{m}(x)
\end{array}\right) - \mxq + e_{1}^\T\left(N_x^\T W_x  N_x\right)^{-1} N_x^\T W_x  \left\{\frac{1}{2} Q_{m}(x)+R_{m}(x) \right\} \\
&=&e_{1}^\T\left(N_x^\T W_x  N_x\right)^{-1} N_x^\T W_x  \left\{\frac{1}{2} Q_{m}(x)+R_{m}(x) \right\}.
\end{eqnarray*}

Using standard results from density estimation \citep[e.g.,][]{Ruppert1994}, if $x \notin \{X_1, \dots, X_n\}$,
\begin{equation}\label{eqn::in1}
n^{-1} \sum_{i=1}^{n} K_{H}\left(X_{i}-x\right)=f(x)+o_{P}(1),
\end{equation}
\begin{equation}\label{eqn::in2}
n^{-1} \sum_{i=1}^{n} K_{H}\left(X_{i}-x\right)\left(X_{i}-x\right)=\mu_{2}(K) H D_{f}(x)+o_{P}(H \bm{1}),
\end{equation}
\begin{equation}\label{eqn::in3}
n^{-1} \sum_{i=1}^{n} K_{H}\left(X_{i}-x\right)\left(X_{i}-x\right)\left(X_{i}-x\right)^\T=\mu_{2}(K) f(x) H+o_{P}(H).
\end{equation}
If $x = X_q  \in \{X_1, \dots, X_n\}$ for some $q=1,\ldots,n$,
\begin{eqnarray}\label{eqn::notin1}
\nonumber n^{-1} \sum_{i=1}^{n} K_{H}\left(X_{i}-x\right)
&=& \frac{n-1}{n} \frac{1}{n-1} \sum_{i\neq q} K_{H}\left(X_{i}-x\right) + \frac{1}{n} K_H(0)\\
& &= f(x)+o_{P}(1),
\end{eqnarray}
\begin{eqnarray}\label{eqn::notin2}
\nonumber n^{-1} \sum_{i=1}^{n} K_{H}\left(X_{i}-x\right)\left(X_{i}-x\right)
\nonumber &=& \frac{n-1}{n} \frac{1}{n-1} \sum_{i\neq q}  K_{H}\left(X_{i}-x\right) \\
&=& \mu_{2}(K) H D_{f}(x)+o_{P}(H \bm{1}),
\end{eqnarray}
\begin{eqnarray}\label{eqn::notin3}
\nonumber & & n^{-1} \sum_{i=1}^{n} K_{H}\left(X_{i}-x\right)\left(X_{i}-x\right)\left(X_{i}-x\right)^\T \\
\nonumber &=&\frac{n-1}{n} \frac{1}{n-1} \sum_{i\neq q}K_{H}\left(X_{i}-x\right)\left(X_{i}-x\right)\left(X_{i}-x\right)^\T \\
&=&\mu_{2}(K) f(x) H+o_{P}(H).
\end{eqnarray}
It follows from above that
{\small \begin{eqnarray}\label{eqn::C4}
\nonumber & & \big( n^{-1} N_x^\T W_x  N_x\big)^{-1} \\
\nonumber &=& \left[\begin{array}{cc}
n^{-1} \sum_{i=1}^{n} K_{H}\left(X_{i}-x\right) & n^{-1} \sum_{i=1}^{n} K_{H}\left(X_{i}-x\right)\left(X_{i}-x\right)^\T \\
n^{-1} \sum_{i=1}^{n} K_{H}\left(X_{i}-x\right)\left(X_{i}-x\right) & n^{-1} \sum_{i=1}^{n} K_{H}\left(X_{i}-x\right)\left(X_{i}-x\right)\left(X_{i}-x\right)^\T
\end{array}\right]^{-1} \\
&=& \left(\begin{array}{cc}
           A_{11} & A_{12} \\
           A_{21} & A_{22}
         \end{array}\right)^{-1} = \left(\begin{array}{cc}
           A^{11} & A^{12} \\
           A^{21} & A^{22}
         \end{array}\right),
\end{eqnarray}}
where we use the following definitions:
$$A_{11} = n^{-1} \sum_{i=1}^{n} K_{H}\left(X_{i}-x\right), \quad A_{12} =  n^{-1} \sum_{i=1}^{n} K_{H}\left(X_{i}-x\right)\left(X_{i}-x\right)^\T, $$
$$A_{21} =  n^{-1} \sum_{i=1}^{n} K_{H}\left(X_{i}-x\right)\left(X_{i}-x\right), \quad A_{22} = n^{-1} \sum_{i=1}^{n} K_{H}\left(X_{i}-x\right)\left(X_{i}-x\right)\left(X_{i}-x\right)^\T,$$
and $A^{11}, A^{12}, A^{21}, A^{22}$ are the corresponding block matrices (vectors) of the inverse matrix $\left(\begin{array}{cc}
           A_{11} & A_{12} \\
           A_{21} & A_{22}
         \end{array}\right)^{-1}$.
And we have
$$
 A^{11}=\left(A_{11}-A_{12} A_{22}^{-1} A_{21}\right)^{-1}, \quad A^{22}=\left(A_{22}-A_{21} A_{11}^{-1} A_{12}\right)^{-1},
$$
$$
A^{12}=-A_{11}^{-1} A_{12} A^{22}, \quad A^{21}=-A_{22}^{-1} A_{21} A^{11}.
$$
Simple algebra gives
{\small\begin{eqnarray*}
 A_{12} A_{22}^{-1} A_{21}
&=& \left\{\mu_2(k) D_f(x)^{\T} H+o_P\left(\bm{1}^{\T} H\right)\right\} \cdot \left\{\mu _ { 2 } ( k ) f(x)H + o _ { p } ( H ) \right\} ^ {-1} \cdot \left\{\mu_2(k) H D_f(x)+o_P(H\bm{1})\right\} \\
&=& \left\{\mu _ {2} ( k ) D_f( x) ^ {\T} H + o_P ( \bm{1}^{\T} H )\right\}\cdot \left\{\mu_{2}(k)^{-1} f(x)^{-1} H^{-1}+o_P\left(H^{-1}\right) \right\}  \\
& & \cdot \left\{\mu_2(k) H D_f(x)+o_P(H\bm{1})\right\} \\
&=&\left\{f(x)^{-1} D_f (x)^{\T}+o_P\left(D_f(x)^{\T}+f(x)^{-1} \bm{1}^{\T}+\bm{1}^{\T}\right)\right\} \cdot \left\{\mu_2(k) H D_f( x)+o_P(H \bm{1})\right\} \\
&=&\mu_2(k) D_f(x)^{\T} H D_f(x)+o_P\{\textnormal{tr}(H)\}, \\
 A^{11}
&=&\left[f(x)+o_P(1)-\mu_2(k) D_f(x)^{\T} H D_f(x)+o_P\{tr(H)\}\right]^{-1} \\
&=&\{f(x)+o_P(1)\}^{-1}, \quad \text{as each entry of } H \to 0 \\
&=&f(x)^{-1}+o_P(1),\\
A_{21} A_{11}^{-1} A_{12} &=&\left\{\mu_2(k) H D_f(x)+o_P(H\bm{1})\right\}\cdot \{f(x)+o_P(1)\}^{-1} \cdot \left\{\mu_2(k) D_f(x)^{\T} H+o_P\left(\bm{1}^{\T} H\right)\right\} \\
&=&\left\{f(x) \mu_2(k) H D_f(x)+o_P\left(\mu_2(k) H D_f(x)+f(x) H \bm{1}+H\bm{1}\right)\right\}\\
&& \cdot\left\{\mu_2(k) D_f(x)^{\T} H+o_P\left(\bm{1}^{\T}H\right)\right\}\\
&=&f(x) \mu_2(k)^2 H D_f(x) D_f(x)^{\T} H+ o_P\left(H^2\right), \\
A^{22} &=&\left\{\mu_2(k) f(x) H+o_P(H)-f(x) \mu_2(k)^2 H f(x) D_f(x)^{T} H+o_P(H^2) \right\}^{-1} \\
&=&\left\{\mu_2(k) f(x) H+o_P(H)\right\}^{-1} \\
&=&\left\{\mu_2(k) f(x) H\right\}^{-1}+o_P\left(H^{-1}\right), \\
A^{12}&=&-\left\{f(x)^{-1}+o_P(1)\right\}\left\{\mu_2(k) D_f(x)^{\T} H+o_P(\bm{1}^\T H)\right\} \cdot \big[\left\{\mu_2(k) f(x) H\right\}^{-1}+o_P\left(H^{-1}\right)\big] \\
&=&-\left\{\mu_2(k) f(x)^{-1} D_f(x)^{\T} H+o_P\left(D_f(x)^{\T} H\right)\right\} \cdot \left\{\mu_2(k)^{-1} f(x)^{-1} H^{-1}+o_P\left(H^{-1}\right)\right\} \\
&=&-D_f(x)^{\T} f(x)^{-2} + o_P(\bm{1}^\T),
\end{eqnarray*}}
and by similar deduction, we have $A^{21} = D_f(x)f(x)^{-2} + o_P(\bm{1})$. Therefore,
\begin{equation*} \big(n^{-1} N_x^\T W_x  N_x\big)^{-1} = \left[\begin{array}{cc}
f(x)^{-1} + o_P(1) & -D_f(x)^\T f(x)^{-2} + o_P(\bm{1^\T}) \\
-D_f(x) f(x)^{-2} + o_P(\bm{1}) & \{\mu_2(K)f(x)H\}^{-1} + o_P(H^{-1}),
\end{array}\right],
\end{equation*}
\begin{equation*}
n^{-1} N_x^{\T} W_x  Q_{m}(x) = \left[\begin{array}{c}
n^{-1} \sum_{i=1}^{n} K_{H}\left(X_{i}-x\right)\left(X_{i}-x\right)^{\T} \mathcal{H}_{m}(x)\left(X_{i}-x\right) \\
n^{-1} \sum_{i=1}^{n}\left\{K_{H}\left(X_{i}-x\right)\left(X_{i}-x\right)^{\T} \mathcal{H}_{m}(x)\left(X_{i}-x\right)\right\}\left(X_{i}-x\right)
\end{array}\right],
\end{equation*}
\begin{eqnarray*}
& & n^{-1}  \sum_{i=1}^{n} K_{H}\left(X_{i}-x\right)\left\{\left(X_{i}-x\right)^{\T} \mathcal{H}_{m}(x)\left(X_{i}-x\right)\right\}\left(X_{i}-x\right) \\
&=& \int K(u)\left\{\left(H^{1 / 2} u\right)^{T} \mathcal{H}_{m}(x)\left(H^{1 / 2} u\right)\right\}\left(H^{1 / 2} u\right) f\left(x+H^{1 / 2} u\right) d u +o_{P}\left(H^{3 / 2} \bm{1}\right) \\
&=& O_{P}\left(H^{3 / 2} \bm{1}\right).
\end{eqnarray*}

Thus,
\begin{equation}
\begin{aligned}
&E\left\{\mhxq \mid X_{1}, \ldots, X_{n}\right\}-\mxq \\
=& \frac{1}{2} e_{1}^\T\left(N_x^\T W_x  N_x\right)^{-1} N_x^\T W_x  Q_{m}(x)+o_P\{\textnormal{tr}(H)\}\\
=& \frac{1}{2} f(x)^{-1} E\Big\{n^{-1} \sum_{i=1}^{n} K_{H}\left(X_{i}-x\right)\left(X_{i}-x\right)^{\T} \mathcal{H}_{m}(x)\left(X_{i}-x\right)\Big\} +o_{P}\{\textnormal{tr}(H)\} \\
=& \frac{1}{2} f(x)^{-1}\left\{\int K(u)\left(H^{1 / 2} u\right)^{\T} \mathcal{H}_{m}(x)\left(H^{1 / 2} u\right) f\left(x+H^{1 / 2} u\right) d u\right\} +o_{P}\{\textnormal{tr}(H)\} \\
=& \frac{1}{2} \textnormal{tr}\left\{H^{1 / 2} \mathcal{H}_{m}(x) H^{1 / 2} \int K(u) u u^{T} d u\right\}+o_{P}\{\textnormal{tr}(H)\} \\
=& \frac{1}{2} \mu_{2}(K) \textnormal{tr}\left\{H \mathcal{H}_{m}(x)\right\}+o_{P}\{\textnormal{tr}(H)\}. \nonumber
\end{aligned}
\end{equation}

For the variance, we have
\begin{equation}
\operatorname{Var} \left\{\mhxq \mid X_{1}, \ldots, X_{n}\right\} =e_{1}^{\T}\left(N_x^{\T} W_x  N_x\right)^{-1} N_x^{\T} W_x  V W_x  N_x\left(N_x^{\T} W_x  N_x\right)^{-1} e_{1}. \nonumber
\end{equation}
The upper-left entry of $n^{-1}N_x^\T W_x V W_x N_x$ is
\begin{eqnarray*}
\nonumber & & n^{-1} \sum_{i=1}^{n} K_{H}\left(X_{i}-x\right)^{2} \nu\left(X_{i}\right) \\
\nonumber &=&|H|^{-1 / 2} \int K^{2}(u) \nu\left(x+H^{1 / 2} u\right) f\left(x+H^{1 / 2} u\right) d u\left\{1+o_{P}(1)\right\} \\
&=&|H|^{-1 / 2} R(K) \nu(x) f(x)\left\{1+o_{P}(1)\right\},
\end{eqnarray*}
where $R(K) = \int K(u)^2 du$. The upper-right block of $n^{-1}N_x^\T W_x V W_x N_x$ is
\begin{eqnarray*}
& & n^{-1} \sum_{i=1}^{n} K_{H}\left(X_{i}-x\right)^{2}\left(X_{i}-x\right)^{\T} \nu\left(X_{i}\right) \\
&=& |H|^{-1 / 2} \int K^{2}(u) u^{\T} H^{1 / 2} \nu\left(x+H^{1 / 2} u\right) f\left(x+H^{1 / 2} u\right) d u\left\{1+o_{P}(1)\right\} \\
&=& O_{P}\left(|H|^{1 / 2}\right),
\end{eqnarray*}
and the lower-right block is
\begin{eqnarray*}
& & n^{-1} \sum_{i=1}^{n} K_{H}\left(X_{i}-x\right)^{2}\left(X_{i}-x\right)\left(X_{i}-x\right)^{T} \nu\left(X_{i}\right) \\
&=& |H|^{-1 / 2} H^{1 / 2}\left\{\int K^{2}(u) u u^{T} d u\right\} H^{1 / 2} \nu(x) f(x)+o_{P}\left(|H|^{-1 / 2} H\right).
\end{eqnarray*}
Using equation (\ref{eqn::C4}) again and the above equations, we have
\begin{equation}
\operatorname{Var} \left\{\mhxq \mid X_{1}, \ldots, X_{n}\right\}=n^{-1}|H|^{-1 / 2}\{R(K) \nu(x) / f(x)\}\left\{1+o_{P}(1)\right\}.
\end{equation}
Then
\begin{equation*}
\begin{aligned}
& E\left[\left\{\hat{m}(x) - m(x)\right\}^2|X_1,\dots,X_n\right] \\
=& n^{-1}|H|^{-1 / 2}\{R(K) \nu(x) / f(x)\} + \frac{1}{4} \mu_{2}(K)^2 \textnormal{tr}^2\left\{H \mathcal{H}_{m}(x)\right\}+o_{P}\{n^{-1}|H|^{-1/2} + \textnormal{tr}^2(H)\} \\
\stackrel{P}{\rightarrow}& 0,
\end{aligned}
\end{equation*}
under Assumption~(iii).
\end{proof}

\subsection{Proof of Theorem~\ref{Theo5}}
\begin{proof}
\begin{eqnarray}
\label{eqn::theo1}
  \nonumber & & \quad \sqrt{n}  \Big[ \big\{ \barhathat - \barhat \big\}  -  \big\{ \barhathac - \barhac \big\}  \Big]  \\
  \nonumber &=& \sqrt{n} \left[\frac{1}{\nkt} \sumik A_i \{ \hatha - \ha\} - \frac{1}{\nkc} \sumik (1-A_i) \{\hatha - \ha\}\right] \\
   &=& \frac{1}{\sqrt{n}\pnk \pink (1-\pink)} \sumik (A_i - \pink) \{\hatha - \ha\}.
\end{eqnarray}
Decomposing $\hatha - \ha$ into the variance term and the bias term, conditional on $X_1, \dots, X_n$, and by Taylor's expansion, we have
\begin{eqnarray}
\label{eqn::theo2}
\nonumber  & & \hatha - \ha \\
\nonumber  &=& e_1^{\T} (N_i^{\T} W_i N_i)^{-1} N_i^{\T} W_i (Y(a) - h_{[k]}(a)) + e_1^{\T} (N_i^{\T} W_i N_i)^{-1} N_i^{\T} W_i h_{[k]}(a) - \ha  \\
\nonumber &=& e_1^{\T} (N_i^{\T} W_i N_i)^{-1} N_i^{\T} W_i \varepsilon(\nu) + e_1^{\T} (N_i^{\T} W_i N_i)^{-1} N_i^{\T} W_i \\
\nonumber &&  \cdot \left\{ N_i \left[\begin{array}{c}
                                                            \ha \\
                                                            D_{h_{[k]}}(X_i,a)
                                                          \end{array}\right]+ \frac{1}{2}Q_{h_{[k]}}(X_i,a) + R_{h_{[k]}}(X_i,a)\right\} - \ha  \\
&=& e_1^{\T} (N_i^{\T} W_i N_i)^{-1} N_i^{\T} W_i \varepsilon(\nu) + \frac{1}{2} e_1^{\T} (N_i^{\T} W_i N_i)^{-1} N_i^{\T} W_i Q_{h_{[k]}}(X_i,a) + o_P(\textnormal{tr}(H)),
\end{eqnarray}
where
$$ W_i = diag(K_H(X_i - X_1), \dots, K_H(X_i - X_n)),$$
$$N_i = \begin{pmatrix}
        1 & (X_1 - X_i)^{\T} \\
        \vdots & \vdots \\
        1 & (X_n - X_i)^{\T}
      \end{pmatrix},$$
$$Y(a) = (Y_1(a), \dots, Y_n(a))^\T, \quad h_{[k]}(a) = (h_{[k]}(X_1,a),\dots,h_{[k]}(X_n,a))^{\T},$$
$$ \varepsilon(\nu) = (\nu(X_1)^{1/2}\varepsilon_1, \dots, \nu(X_n)^{1/2}\varepsilon_n)^{\T},$$
$$ Q_{h_{[k]}}(X_i,a) = \Big((X_1 - X_i)^\T \mathcal{H}_{h_{[k]}}(X_i,a)(X_1 - X_i), \dots, (X_n - X_i)^\T \mathcal{H}_{h_{[k]}}(X_i,a)(X_n - X_i)\Big)^T,$$
and $R_{h_{[k]}}(X_i,a)$ is the vector of Taylor series remainder terms. Here $\mathcal{H}_{h_{[k]}}(X_i,a)$ denotes the $d \times d$ Hessian matrix of $h_{[k]}(\cdot,a)$ evaluated at $X_i$.

Following the proof of the ordinary least squares-adjusted estimators, we split the variance term and the bias term into a product of an $O_P(1)$ term and an $o_P(1)$ term, respectively. For the variance term, by the derivation of Lemma~\ref{lem::ks} and equations (\ref{eqn::in1})--(\ref{eqn::notin3}), we have

\begin{eqnarray}
\label{eqn::theo3}
  \nonumber & & e_1^{\T} (N_i^{\T} W_i N_i)^{-1} N_i^{\T} W_i \varepsilon(\nu) \\
\nonumber   &=& e_1^{\T}  (n^{-1} N_i^{\T} W_i N_i )^{-1} \{n^{-1}N_i^{\T} W_i \varepsilon(\nu)\} \\
\nonumber   &=&  e_1^\T \left[\begin{array}{cc}
 f(X_i)^{-1} + o_P(1) & - D_f(X_i)^\T f(X_i)^{-2} + o_P(\bm{1^\T}) \\
 -D_f(X_i) f(X_i)^{-2} + o_P(\bm{1}) &  \{\mu_2(K)f(X_i)H\}^{-1} + o_P(H^{-1})
\end{array}\right] \\
\nonumber && \cdot \left[\begin{array}{c}
 n^{-1} \sum_{j=1}^{n} K_{H}(X_{j}-X_i) \nu^{1/2}(X_j) \varepsilon_j \\
 n^{-1} \sum_{j=1}^{n} K_{H}(X_{j}-X_i)(X_j - X_i) \nu^{1/2}(X_j) \varepsilon_j
\end{array}\right]\\
\nonumber &=& f(X_i)^{-1}n^{-1} \sum_{j=1}^{n} K_{H}(X_{j}-X_i) \nu^{1/2}(X_j) \varepsilon_j \\
\nonumber && - f(X_i)^{-2}D_f(X_i)^\T n^{-1} \sum_{j=1}^{n} K_{H}(X_{j}-X_i)(X_j - X_i) \nu^{1/2}(X_j) \varepsilon_j + o_P(1).
\end{eqnarray}

Denote
\begin{eqnarray*}
  \xi_1 &=& n^{-1/2}\sumik (A_i - \pink) f(X_i)^{-1} n^{-1} \sum_{j=1}^{n} K_{H}(X_{j}-X_i) \nu^{1/2}(X_j) \varepsilon_j \\
  &=& n^{-1} \sum_{j=1}^{n} \big\{n^{-1/2} \sumik (A_i - \pink) f(X_i)^{-1} K_{H}(X_{j}-X_i)\big\} \nu^{1/2}(X_j) \varepsilon_j,
\end{eqnarray*}
where $f(X_i) > 0, \ K_H(X_j - X_i) > 0,\ \nu(X_j)>0$, $A_i - \pink$ and $f(X_i)$ are bounded. Then
$$E(\xi_1) = E\{E(\xi_1|X_1,\dots,X_n)\} = 0,$$
{\small\begin{eqnarray*}
\Var(\xi_1) &=& E\{\Var(\xi_1|X_1,\dots,X_n,A^{(n)})\} + \Var\{E(\xi_1|X_1,\dots,X_n,A^{(n)})\} \\
   &=& E\{\Var(\xi_1|X_1,\dots,X_n,A^{(n)})\}\\
   &=& E \Big(\Var \big[n^{-1} \sum_{j=1}^{n} \big\{n^{-1/2} \sumik (A_i - \pink) f(X_i)^{-1} K_{H}(X_{j}-X_i)\Big\} \\
   & & \cdot \nu^{1/2}(X_j) \varepsilon_j|X_1,\dots,X_n, A^{(n)}\big]\Big) \\
   &=& E\Big[n^{-2} \sumj n^{-1} \Big\{\sumik (A_i - \pink) f(X_i)^{-1} K_H(X_j-X_i)\Big\}^2 \nu(X_j)\Big]\\
   &=& E\Big(n^{-1} \sumj n^{-1}|H|^{-1} \cdot n^{-1} \Big[\sumik (A_i - \pink) f(X_i)^{-1} K\{H^{-1/2}(X_j-X_i)\}\Big]^2 \nu(X_j)\Big).
\end{eqnarray*}}
There exist a constant $C_{\xi}$ such that $0 \leq n^{-1} [\sumik (A_i - \pink) f(X_i)^{-1} K\{H^{-1/2}(X_j-X_i)\}]^2 \leq C_{\xi}$, and $n^{-1}|H|^{-1} \to 0$, so $\Var(\xi_1) \to 0$ as $n \to \infty$. From above, we have
$$\frac{1}{\sqrt{n}\pnk \pink (1-\pink)} \sumik (A_i - \pink)f(X_i)^{-1} n^{-1} \sum_{j=1}^{n} K_{H}(X_{j}-X_i) \nu^{1/2}(X_j) \varepsilon_j  \stackrel{P}{\rightarrow} 0.$$

By similar deduction, we have
$$ C_n^{-1} \sumik (A_i - \pink) f(X_i)^{-2}D_f(X_i)^\T n^{-1} \sum_{j=1}^{n}K_{H}\left(X_{j}-X_i\right)(X_{j}-X_i)\nu^{1/2}(X_j) \varepsilon_j  \stackrel{P}{\rightarrow} 0,$$
where $C_n = \sqrt{n}\pnk \pink (1-\pink).$
Moreover, $n^{-1/2} \sumik (A_i - \pink) = O_P(1)$, so $n^{-1/2} \sumik (A_i - \pink)\cdot o_P(1) = o_P(1)$. Then
$$\frac{1}{\sqrt{n}\pnk \pink (1-\pink)} \sumik (A_i - \pink) e_1^{\T} (N_i^{\T} W_i N_i)^{-1} N_i^{\T} W_i \varepsilon(\nu)\stackrel{P}{\rightarrow} 0.$$

For the second term of equation~(\ref{eqn::theo2}), i.e., the bias term, we have
$$\begin{aligned}
   &  e_1^{\T} (N_i^{\T} W_i N_i)^{-1} N_i^{\T} W_i Q_{h_{[k]}}(X_i,a) \\
   =&  e_1^\T (n^{-1} N_i^{\T} W_i N_i)^{-1} \cdot (n^{-1} N_i^{\T} W_i Q_{h_{[k]}}(X_i,a))\\
   =&  e_1^\T \left[\begin{array}{cc}
 f(X_i)^{-1} + o_P(1) & - D_f(X_i)^\T f(X_i)^{-2} + o_P(\bm{1^\T}) \\
 -D_f(X_i) f(X_i)^{-2} + o_P(\bm{1}) &  \{\mu_2(K)f(X_i)H\}^{-1} + o_P(H^{-1})
\end{array}\right] \\
& \cdot {\left[\begin{array}{c}
n^{-1} \sumj K_{H}\left(X_{j}-X_i\right)\left(X_{j}-X_i\right)^{\T} \mathcal{H}_{h_{[k]}}(X_i,a)\left(X_{j}-X_i\right) \\
n^{-1} \sumj \left\{K_{H}\left(X_{j}-X_i\right)\left(X_{j}-X_i\right)^{\T} \mathcal{H}_{h_{[k]}}(X_i,a)\left(X_{j}-X_i\right)\right\}\left(X_{j}-X_i\right)
\end{array}\right] } \\
=& n^{-1}f(X_i)^{-1}\sumj K_{H}\left(X_{j}-X_i\right)\left(X_{j}-X_i\right)^{\T} \mathcal{H}_{h_{[k]}}(X_i,a)\left(X_{j}-X_i\right) + O_P(D_f(X_i)H^{3/2}\bm{1}) \\
=& f(X_i)^{-1} \left[\left\{\int K(u)\left(H^{1 / 2} u\right)^{T} \mathcal{H}_{h_{[k]}}(X_i,a)\left(H^{1 / 2} u\right) f\left(X_i+H^{1 / 2} u\right) d u\right\} +o_{P}\{\textnormal{tr}(H)\}\right]\\
=& f(X_i)^{-1} \left[ \textnormal{tr}\left\{H^{1 / 2} \mathcal{H}_{h_{[k]}}(X_i,a) H^{1 / 2} \int K(u) u u^{T} d u\right\}+o_{P}\{\textnormal{tr}(H)\}\right] \\
=& f(X_i)^{-1} \mu_{2}(K) \textnormal{tr}\left\{H \mathcal{H}_{h_{[k]}}(X_i,a)\right\} + o_{P}\{\textnormal{tr}(H)\}.
\end{aligned}$$

Let $C_H = \underset{i,j \in \{1,\dots,d\}}{\max}H_{ij}$. Because each entry of $H$ tends to $0$, $C_H$ tends to $0$ as well. Recall that all second-order derivatives of $h_{[k]}(\cdot,a)$ are continuous, and the density function of $X_i$'s has a compact support set, so the elements of $\mathcal{H}_{h_{[k]}}(X_i,a)$ are also bounded. Then there exist a constant $C_{\mathcal{H}}$ such that
$$|\textnormal{tr}\left\{H \mathcal{H}_{h_{[k]}}(X_i,a)\right\}| \leq d^2 C_H C_{\mathcal{H}}.$$
Therefore,
\begin{eqnarray*}
   && n^{-1/2} \sumik A_i f(X_i)^{-1}e_1^{\T} (N_i^{\T} W_i N_i)^{-1} N_i^{\T} W_i Q_{h_{[k]}}(X_i,a) \\
   &=& \mu_2(K) n^{-1/2} \sumik A_i f(X_i)^{-1} \left[\textnormal{tr}\left\{H \mathcal{H}_{h_{[k]}}(X_i,a)\right\} + o_{P}\{\textnormal{tr}(H)\}\right],
\end{eqnarray*}
where $n^{-1/2} \sumik A_i f(X_i)^{-1} \textnormal{tr}\left\{H \mathcal{H}_{h_{[k]}}(X_i,a)\right\} \leq n^{-1/2} \sumik A_i f(X_i)^{-1} d^2 C_H C_{\mathcal{H}}$.
By the central limit theorem in \cite{Bugni2018}, we have $n^{-1/2} \sumik A_i f(X_i)^{-1} = O_P(1)$. Thus,
$$ n^{-1/2} \sumik A_i f(X_i)^{-1}e_1^{\T} (N_i^{\T} W_i N_i)^{-1} N_i^{\T} W_i Q_{h_{[k]}}(X_i,a) = o_P(1).$$

By similar deduction, we have
$$ n^{-1/2} \sumik (1-A_i) f(X_i)^{-1}e_1^{\T} (N_i^{\T} W_i N_i)^{-1} N_i^{\T} W_i Q_{h_{[k]}}(X_i,a) = o_P(1).$$
As a consequence,
$$\frac{1}{\sqrt{n}\pnk \pink (1-\pink)} \sumik (A_i - \pink) \{\hatha - \ha\} = o_P(1),$$
that is, equation~(\ref{eqn::goal1}) holds for local linear kernel, under Assumption~\ref{assum::kernel}.

To prove equation~(\ref{eqn::goal2}), recall that $\{X_i\}_{i=1}^n$ are i.i.d., and conditional on $\{B_1, \dots, B_n\}$, $\{A_1, \dots, A_n\}$ are independent of $\{X_i\}_{i=1}^n$. Similar to the proof of Lemma B.2 in \cite{Bugni2018}, by arranging the order of units with respect to the treatment assignment, we can construct quantities $\tilde{\hat{h}}_{[k]}(X_i,a)$'s such that $\tilde{\hat{h}}_{[k]}(X_i,a) \stackrel{d}{=} \hatha|B_i=k, a = 0, 1$, and $\tilde{\hat{h}}_{[k]}(X_i,a)$'s don't depend on $B^{(n)}$ and $A^{(n)}$. Then, for $\forall \varepsilon >0$, by Markov inequality, we have
\begin{eqnarray*}
  & & \mathbb{P} \Big(\frac{1}{ \nk } \sumik  \left\{   \tilde{\hat{h}}_{[k]}(X_i,a) - \ha   \right\}^2 > \varepsilon \mid X_1, \dots, X_n  \Big) \\
 &\leq& E\left[ \frac{1}{ \nk } \sumik  \left\{   \tilde{\hat{h}}_{[k]}(X_i,a) - \ha   \right\}^2  \mid X_1, \dots, X_n \right]/\varepsilon \\
   &\xrightarrow{P}& 0,
\end{eqnarray*}
by lemma~\ref{lem::ks}. Thus,
$$
\mathbb{P} \Big(\frac{1}{ \nk } \sumik  \left\{   \tilde{\hat{h}}_{[k]}(X_i,a) - \ha   \right\}^2 > \varepsilon \Big) \xrightarrow{P} 0.
$$
Hence, $ \nk^{-1} \sumik  \big\{   \hatha - \ha   \big\}^2 = o_P(1)$, i.e., equation~(\ref{eqn::goal2}) holds.
\end{proof}

\section{Proof of Theorem~\ref{theo::ss}}
\begin{proof}
Let $\nkm = \sumkm 1$ denote the number of units in stratum $k$ and fold $m$, $\nkam = \sumkm \mathds{1}_{A_i = a}, a = 0, 1,$ denote the number of units in stratum $k$ and fold $m$ with treatment $a$. For simplicity, suppose each fold has the same number of units $\nm$, then $n = M \nm$. Denote $\pnkm = \nkm/\nm = (M\nkm)/n, \pinkm = \nktm/\nkm$, then ${\nkm}/{\nk} \stackrel{P}{\to} 1/M,$
\begin{eqnarray}
& & \sumk \pnk \YkThat -  \frac{1}{M}\summ \sumk \pnkm \frac{1}{\nktm} \sumkm A_i Y_i \nonumber \\
&=& \sumk \pnk \summ \frac{\nktm}{\nkt} \cdot \frac{1}{\nktm} \sumkm A_i Y_i - \frac{1}{M}\summ \sumk \pnkm \cdot \frac{1}{\nktm} \sumkm A_i Y_i \nonumber \\
&=& \summ \sumk \pnk \frac{\nktm}{\nkt} \cdot \frac{1}{\nktm} \sumkm A_i Y_i - \summ \sumk \pnk \frac{\nkm}{\nk} \cdot \frac{1}{\nktm} \sumkm A_i Y_i \nonumber \\
&=& \summ \sumk \pnk (\frac{\nktm}{\nkt} - \frac{\nkm}{\nk})\cdot \frac{1}{\nktm} \sumkm A_i Y_i. \label{eqn::diff1}
\end{eqnarray}
By similar deduction, we have
\begin{eqnarray}
& & \sumk \pnk \frac{1}{\nkt} A_i \hts -  \frac{1}{M}\summ \sumk \pnkm \frac{1}{\nktm} \sumkm A_i \hathtms \nonumber \\
&=& \sumk \pnk \summ \frac{\nktm}{\nkt} \cdot \frac{1}{\nktm} \sumkm A_i \hts \nonumber \\
& & - \frac{1}{M}\summ \sumk \pnkm \frac{1}{\nktm} \sumkm A_i \hathtms \nonumber \\
&=& \summ \sumk \pnk \frac{\nktm}{\nkt} \cdot \frac{1}{\nktm} \sumkm A_i \hts  \nonumber\\
& & - \summ \sumk \pnk \frac{\nkm}{\nk} \cdot \frac{1}{\nktm} \sumkm A_i \hathtms  \nonumber \\
&=& \summ \sumk \pnk (\frac{\nktm}{\nkt} - \frac{\nkm}{\nk})\cdot \frac{1}{\nktm} \sumkm A_i \hts \nonumber \\
& & + \summ \sumk \pnk \frac{\nkm}{\nk} \cdot \frac{1}{\nktm} \sumkm A_i \{\hts - \hathtms\},\label{eqn::diff2}
\end{eqnarray}
and
\begin{eqnarray}
& & \frac{1}{M}\summ \sumk \pnkm \frac{1}{\nktm} \sumkm \pinkm \hathtms \nonumber \\
& & -  \sumk \pnk \frac{1}{\nkt} \sumik \pink \hts \nonumber\\
&=& \summ \sumk \pnk \frac{\nkm}{\nk} \frac{1}{\nktm} \sumkm \pinkm \hathtms \nonumber \\
& & - \sumk \pnk \summ \frac{\nktm}{\nkt} \frac{1}{\nktm} \sumkm \frac{\pink}{\pinkm}\pinkm \hts \nonumber \\
&=& \summ \sumk \pnk \frac{\nkm}{\nk} \frac{1}{\nktm} \sumkm \pinkm \hathtms \nonumber \\
& & - \summ \sumk \pink \frac{\nkm}{\nk} \frac{1}{\nktm} \sumkm \pinkm \hts \nonumber \\
&=& \summ \sumk \pnk \frac{\nkm \pinkm}{\nk} \frac{1}{\nktm} \sumkm \{\hathtms - \hts\} \label{eqn::diff3}
\end{eqnarray}
Because the folds are mutually exclusive and the fold partition process is independent of covariates, stratum, treatments, and outcomes, we have $\pinkm \stackrel{P}{\to} \pi$ as $n \to \infty$. Moreover, $\hts = E(Y_i|B_i = k,A_i=1)$ implies $ ({1}/{\nktm}) \sumkm A_i \{Y_i - \hts\} = O_p(1/\sqrt{n})$. 
Together with $\pink \stackrel{P}{\to} \pi$ and Equations~\eqref{eqn::diff1},~\eqref{eqn::diff2},~\eqref{eqn::diff3}, we have

\begin{eqnarray*}
  & & \sumk \pnk \YkThat -  \sumk \pnk \frac{1}{\nkt} (A_i-\pink) \hts\\
  & & - \Big\{ \frac{1}{M}\summ \sumk \pnkm \frac{1}{\nktm} \sumkm A_i Y_i \\
  & & - \frac{1}{M}\summ \sumk \pnkm \frac{1}{\nktm} \sumkm (A_i-\pinkm) \hathtms \Big\} \\
  &=& \summ \sumk \pnk (\frac{\nktm}{\nkt} - \frac{\nkm}{\nk})\cdot \frac{1}{\nktm} \sumkm A_i Y_i \\
  & & - \summ \sumk \pnk (\frac{\nktm}{\nkt} - \frac{\nkm}{\nk})\cdot \frac{1}{\nktm} \sumkm A_i \hts \\
  & & + \summ \sumk \pnk \frac{\nkm}{\nk} \cdot \frac{1}{\nktm} \sumkm A_i \{\hts - \hathtms\} \\
  & & + \summ \sumk \pnk \frac{\nkm \pinkm}{\nk} \frac{1}{\nktm} \sumkm \{\hathtms - \hts\} \\
  &=& \summ \sumk \pnk (\frac{\nktm}{\nkt} - \frac{\nkm}{\nk})\cdot \frac{1}{\nktm} \sumkm A_i \{Y_i - \hts\} \\
  & & + \summ \sumk \pnk \frac{\nkm}{\nk} \cdot \frac{1}{\nktm} \sumkm (A_i-\pinkm) \{\hts - \hathtms\} \\
  &=& \summ \sumk \pnk \frac{\nkm(\pinkm - \pink)}{\nkt}\cdot \frac{1}{\nktm} \sumkm A_i \{Y_i - \hts\} \\
  & & + \summ \sumk \pnk \frac{\nkm}{\nk} \cdot \frac{1}{\nktm} \sumkm (A_i-\pinkm) \{\hts - \hathtms\} \\
  &=& \summ \sumk \pnk \frac{\nkm(\pinkm - \pi)}{\nkt}\cdot \frac{1}{\nktm} \sumkm A_i \{Y_i - \hts\} \\
  & & + \summ \sumk \pnk \frac{\nkm(\pi - \pink)}{\nkt}\cdot \frac{1}{\nktm} \sumkm A_i \{Y_i - \hts\} \\
  & & + \summ \sumk \pnk \frac{\nkm}{\nk} \cdot \frac{1}{\nktm} \sumkm (A_i-\pinkm) \{\hts - \hathtms\} \\
  &=& \summ \sumk \pnk \frac{\nkm}{\nk} \cdot \frac{1}{\nktm} \sumkm (A_i-\pinkm) \{\hts - \hathtms\} + o_P({1}/{\sqrt{n}}).
\end{eqnarray*}
By similar deduction, we can establish the corresponding results for the control group:
\begin{eqnarray*}
  & & \sumk \pnk \YkChat +  \sumk \pnk \frac{1}{\nkc} (A_i-\pink) \hcs\\
  & & - \Big\{ \frac{1}{M}\summ \sumk \pnkm \frac{1}{\nkcm} \sumkm(1- A_i) Y_i \\
  & & + \frac{1}{M}\summ \sumk \pnkm \frac{1}{\nkcm} \sumkm (A_i-\pinkm) \hathcms \Big\} \\
  &=& \summ \sumk \pnk \frac{\nkm}{\nk} \cdot \frac{1}{\nkcm} \sumkm \{(1-A_i)-(1-\pinkm)\} \\
  & & \cdot \{\hcs - \hathcms\} + o_P({1}/{\sqrt{n}}).
\end{eqnarray*}

Within stratum $k$ and fold $m$, let $\barhathatm$ and $\barhathacm$ respectively denote the sample means of $\hatham$ in the treatment group and control group, and let $\barhatm$ and $\barhacm$ respectively denote the sample means of $\ha$ in the treatment group and control group. Then,
\begin{eqnarray*}
  & & \summ \sumk \pnk \frac{\nkm}{\nk} \cdot \frac{1}{\nktm} \sumkm \big(A_i-\pinkm\big) \big\{\hts - \hathtms\big\} \\
  & & - \summ \sumk \pnk \frac{\nkm}{\nk} \cdot \frac{1}{\nkcm} \sumkm \big\{(1-A_i)-(1-\pinkm)\big\} \big\{\hcs - \hathcms\big\} \\
  &=& - \summ \sumk \pnk \frac{\nkm}{\nk} (1-\pinkm)\Big[\big\{\bar{\hat{h}}_{[k]m1}(\cdot,1) - \bar{h}_{[k]m1}(\cdot,1)\big\} - \big\{\bar{\hat{h}}_{[k]m0}(\cdot,1) - \bar{h}_{[k]m0}(\cdot,1)\big\}\Big] \\
  & & - \summ \sumk \pnk \frac{\nkm}{\nk} \pinkm \Big[\big\{\bar{\hat{h}}_{[k]m1}(\cdot,0) - \bar{h}_{[k]m1}(\cdot,0)\big\} - \big\{\bar{\hat{h}}_{[k]m0}(\cdot,0) - \bar{h}_{[k]m0}(\cdot,0)\big\}\Big],
\end{eqnarray*}

As a consequence, to prove that $\sqrt{n}\tauss$ and $\sqrt{n}\tauor$ have the same asymptotic distribution, we only need to prove the sample splitting version of Assumption~\ref{assum4}, i.e., Assumption~\ref{assum::appendix} below holds.
\begin{Assumption}\label{assum::appendix}
  For $k = 1,\dots, K, m = 1, \dots, M$ and $a=0,1$,
\begin{equation}
\label{eqn::goal1ss}
  \sqrt{n}  \Big[ \big\{ \barhathatm - \barhatm \big\}  -  \big\{ \barhathacm - \barhacm \big\}  \Big] = o_P(1),
\end{equation}
\begin{equation}
\label{eqn::goal2ss}
  \frac{1}{ \nkm } \sumkm  \Big\{   \hatham - \ha   \Big\}^2 = o_P(1).
\end{equation}
\end{Assumption}

For Equation~\eqref{eqn::goal1ss}, we have
\begin{eqnarray}\label{eqn::goal1expand}
   && \sqrt{n}  \Big[ \big\{ \barhathatm - \barhat \big\}  -  \big\{ \barhathacm - \barhac \big\}  \Big] \nonumber \\
   &=& \sqrt{n}\Big(\big[\frac{1}{\nktm}\sumkm A_i\{\hatham - \ha\}\big] \nonumber \\
   & & - \big[\frac{1}{\nkcm}\sumkm (1-A_i) \{\hatham - \ha\}\big]\Big) \nonumber\\
   &=& \frac{\sqrt{n}}{\nkm}\sumkm \big(\frac{A_i}{\pinkm}-\frac{1-A_i}{1-\pinkm}\big)\{\hatham - \ha\} \nonumber \\
   &=& \frac{M}{\pnkm \sqrt{n}} \sumkm \frac{A_i - \pinkm}{\pinkm(1-\pinkm)}\{\hatham - \ha\}.
\end{eqnarray}
For $a = 1$, $$\eqref{eqn::goal1expand} = \frac{M}{\pnkm \sqrt{n}} \sumkm \frac{\mathds{1}_{A_i = 1}-\pinkm}{\pinkm(1-\pinkm)}\{\hatham - \ha\}.$$
For $a=0$,
\begin{eqnarray*}
  \eqref{eqn::goal1expand} &=& -\frac{M}{\pnkm \sqrt{n}} \sumik \frac{(1-A_i)-(1-\pinkm)}{\pinkm(1-\pinkm)}\{\hatham - \ha\} \\
  &=& -\frac{M}{\pnkm \sqrt{n}} \sumkm \frac{\mathds{1}_{A_i = 0}-(1-\pinkm)}{\pinkm(1-\pinkm)}\{\hatham - \ha\}.
\end{eqnarray*}
Let
$$\pinkma = \begin{cases}
                \pinkm, & \mbox{if a = 1}  \\
                1-\pinkm, & \mbox{if a = 0}.
              \end{cases}, \quad \Dika = \hatham - \ha.$$
To prove Equation~\eqref{eqn::goal1ss}, it suffices to show
$$\frac{1}{\sqrt{n}}\sumkm \frac{\mathds{1}_{A_i=a}-\pinkma}{\pinkma}\Dika = o_P(1).$$

Let
$$\Tnkm = \frac{1}{\sqrt{n}}\sumi \frac{\mathds{1}_{A_i=a}-\pinkma}{\pinkma}\{\hatham - \ha\}\cdot \mathds{1}_{i\in [k] \cap I_m}.$$
Denote $A^{(n)} = (A_1, A_2, \dots, A_n)$ and $B^{(n)} = (B_1, B_2, \dots, B_n)$. Suppose that $\Wk'$ and $\Wk''$ are independent copies of $\Wk = \{Y_i(1), Y_i(0), X_i\}_{i\in [k]}, \ k = 1,\dots, K$, such that $(\Wk', \Wk'') \perp \!\!\! \perp (\Wk, A^{(n)})$ and $\Wk' \perp \!\!\! \perp \Wk''$. Let $\Nkam = (\nk, \nkm, \nkam)$. By Lemma~\ref{lem1} and the law of large numbers, we have ${\nkam}/{\nkm}\stackrel{P}{\to} \pi(a)$, where $\pi(1) = \pi$ and $\pi(0) = 1-\pi$.

Denote $\bDika$ as the statistic obtained by replacing the units from $I_m^c$ (the units for estimating $\hat{h}_{[k]m}(\cdot,a)$) by units in $\Wk'$ and replacing the units from $I_m$ by units in $\Wk''$ in $\Dika$, and
$$\bTnkm = \frac{1}{\sqrt{n}}\Big\{\sumnkam \frac{1-\pinkma}{\pinkma} \bDika - \sumnkamc \bDika\Big\},$$
then $\bDika \stackrel{d}{=} \Dika$ and $\bTnkm \stackrel{d}{=} \Tnkm$, conditional on $\{\Wk', A^{(n)}\}$, $\bDika$'s are independent across $i \in [k]$ by the independent assumptions, and it remains to show that $\bTnkm = o_P(1)$. Simple calculation gives
\begin{eqnarray*}
 && E\{\bTnkm \mid \Nkam, \Wk'\} \\
 &=& \frac{1}{\sqrt{n}}\Big[\sumnkam \frac{1-\pinkma}{\pinkma} E\{\bDika|\Nkam, \Wk'\} - \sumnkamc E\{\bDika|\Nkam, \Wk'\}\Big] \\
   &=& \frac{1}{\sqrt{n}} \Big(\frac{\nkam}{\pinkma}-\nkm\Big) E\{\bDika|\Nkam, \Wk'\} \\
   &=& 0.
\end{eqnarray*}
In addition, we have
\begin{eqnarray}
   & & \Var\{\bTnkm \mid \Nkam, \Wk'\} \nonumber \\
   &=& \frac{1}{n}\bigg[\Big\{\frac{1-\pinkma}{\pinkma}\}^2 \nkam Var\{\bDika|\Nkam, \Wk'\Big\} \nonumber \\
   & & + (\nkm - \nkam) Var\{\bDika|\Nkam, \Wk'\}\bigg] \nonumber \\
   &=& \frac{1}{n} \Big(\frac{1-2\pinkma}{\pinkma^2} \nkam + \nkm \Big) Var\{\bDika|\Nkam, \Wk'\} \nonumber \\
   &\leq& \frac{1}{n} \Big(\frac{1-2\pinkma}{\pinkma^2} \nkam + \nkm \Big) E\{\bDika^2|\Nkam, \Wk'\} \nonumber \\
   &=& \Big(\frac{1-2\pinkma}{\pinkma^2}\cdot\frac{\nkam}{n}+\frac{\nkm}{n}\Big)E\{\bDika^2|\Nkam, \Wk'\} \nonumber \\
   &=& \Big(\frac{1-\pinkma}{\pinkma}\cdot \frac{\nkm}{n}\Big) E\{\bDika^2|\Nkam, \Wk'\}. \label{eqn::Var}
\end{eqnarray}

In~\eqref{eqn::Var}, ${(1-\pinkma)}/{\pinkma} = O_P(1), \ 0 < {\nkm}/{n} < 1$. Together with  Assumption~\ref{assum::ss}, we have $$\Var\{\bTnkm \mid \Nkam, \Wk'\} \leq O_P(1) \cdot E\{\bDika^2\mid \Nkam, \Wk'\} = o_p(1).$$
Then for $\forall \varepsilon > 0$, by Markov inequality, we have
\begin{equation}
  P(|\bTnkm| > \varepsilon \mid \Nkam, \Wk') \leq \varepsilon^{-2} E(\bTnkm^2 \mid \Nkam, \Wk') = o_P(1). \nonumber
\end{equation}
Therefore, by the extension of the dominated convergence theorem to convergence in probability, we have
\begin{eqnarray*}
  \lim_{n\to \infty} P(|\bTnkm| > \varepsilon) &=& \lim_{n\to \infty} E(\mathds{1}_{|\bTnkm| > \varepsilon}) \\
   &=& \lim_{n\to \infty} E\{E(\mathds{1}_{|\bTnkm| > \varepsilon}\mid \Nkam, \Wk')\} \\
   &=& \lim_{n\to \infty} E(P(|\bTnkm| > \varepsilon \mid \Nkam, \Wk')) \\
   &=& 0.
\end{eqnarray*}


Moreover, if Assumption~\ref{assum::ss} holds, because $X_i$'s are independent and identically distributed, then equation~\eqref{eqn::goal2ss} holds by the law of large numbers. In conclusion, if Assumption~\ref{assum::ss} holds, then the two equations in Assumption~\ref{assum::appendix} also hold.

Next, we prove the consistency of the variance estimator. Note that
\begin{eqnarray*}
& & \sumk \pnk  \bigg[  \frac{1}{\nkt} \sumik  A_i  \Big\{  \hat r_i(1)  - \frac{1}{\nkt} \sumjk A_j  \hat r_j(1)  \Big\} ^2  \bigg] \\
& & - \frac{1}{M} \summ \sumk \pnkm \frac{1}{\nktm} \sumkm  A_i  \Big\{  \hat r_i(1)  - \frac{1}{\nktm} \sumjkm A_j  \hat r_j(1)  \Big\} ^2  \\
&=& \summ \sumk \pnk (\frac{\nktm}{\nkt} - \frac{\nkm}{\nk}) \frac{1}{\nktm} \\
& & \cdot \sumkm A_i  \bigg[\Big\{  \hat r_i(1)  - \frac{1}{\nkt} \sumjk A_j  \hat r_j(1)  \Big\} ^2  - \Big\{  \hat r_i(1)  - \frac{1}{\nktm} \sumjkm A_j  \hat r_j(1)  \Big\} ^2 \bigg] \\
&=& \summ \sumk \pnk (\frac{\nktm}{\nkt} - \frac{\nkm}{\nk}) \Big\{  \frac{1}{\nktm} \sumjkm A_j  \hat r_j(1)  - \frac{1}{\nkt} \sumjk A_j  \hat r_j(1)  \Big\} \\
& & \cdot \frac{1}{\nktm} \sumkm A_i \Big\{  2 \hat r_i(1) - \frac{1}{\nkt} \sumjk A_j  \hat r_j(1)  - \frac{1}{\nktm} \sumjkm A_j  \hat r_j(1)  \Big\} \\
&=&  \summ \sumk \pnk \frac{\nkm(\pinkm - \pink)}{\nkt} \cdot \bigg[\Big\{\frac{2}{\nktm} \sumkm A_i \hat r_i(1) \Big\}\cdot \Big\{  \frac{1}{\nktm} \sumjkm A_j  \hat r_j(1)  \\
& & - \frac{1}{\nkt} \sumjk A_j  \hat r_j(1)  \Big\} -  \Big\{  \frac{1}{\nktm} \sumjkm A_j  \hat r_j(1)\Big\}^2 + \Big\{\frac{1}{\nkt} \sumjk A_j  \hat r_j(1)\Big\}^2 \bigg] \\
&=& o_P(1).
\end{eqnarray*}

By similar deduction, we have
\begin{eqnarray*}
    && \sumk \pnk  \bigg[  \frac{1}{\nkc} \sumik (1 - A_i ) \Big\{ \hat r_i (0)  -   \frac{1}{\nkc} \sumjk ( 1 - A_j )  \hat r_j (0)  \Big\}^2   \bigg]
     - \frac{1}{M} \summ \sumk \pnkm \\
     && \quad \cdot \frac{1}{\nkcm} \sumkm  (1-A_i)  \Big\{  \hat r_i(0)  - \frac{1}{\nkcm} \sumjkm (1-A_j)  \hat r_j(0)  \Big\} ^2 = o_P(1).
\end{eqnarray*}
Using the techniques developed by \cite{Liu2020Lasso}, Assumption~\ref{assum::appendix} implies that
\begin{eqnarray*}
&&\sumk \pnk  \bigg[  \frac{1}{\nkt} \sumik  A_i  \Big\{  \hat r_i(1) - \frac{1}{\nkt} \sumjk A_j  \hat r_j(1)  \Big\} ^2  \bigg]  \\
&&+ \sumk \pnk  \bigg[  \frac{1}{\nkc} \sumik (1 - A_i ) \Big\{ \hat r_i (0)  -   \frac{1}{\nkc} \sumjk ( 1 - A_j )  \hat r_j (0)  \Big\}^2   \bigg] \\
& = & \varsigma_r^2(\pi) + o_p(1).
\end{eqnarray*}

Let $$n_{m1} = \sum_{i \in I_m} A_i, \ n_{m0} = \sum_{i \in I_m} (1-A_i),\ \pin = \frac{n_1}{n}, \ \pim = \frac{n_{m1}}{n_m},$$
$$ \brtm = \frac{1}{n_{m1}} \sum_{i \in I_m} A_i \hat r_i (1), \ \brcm = \frac{1}{n_{m0}} \sum_{i \in I_m} (1-A_i) \hat r_i (0),$$
$$ \brktm = \frac{1}{\nktm} \sumjkm A_j  \hat r_j(1), \ \brkcm = \frac{1}{\nkcm} \sumjkm (1-A_j)  \hat r_j(0).$$

Since the folds are mutually exclusive and the fold partition process is independent of covariates, stratum, treatments, and outcomes, we can show that $\pin \stackrel{P}{\to} \pi,$ $\pim \stackrel{P}{\to} \pi$,
$
 \brktm \stackrel{P}{\to} E\{r_i(1) \mid B_i = k\}, $ $\brkcm \stackrel{P}{\to} E\{r_i(0) \mid B_i = k\},
$ $\brtm  \stackrel{P}{\to} E\{r_i(1)\}$, and $\brcm  \stackrel{P}{\to} E\{ r_i(0) \}$. Thus,
\begin{eqnarray*}
    &&\frac{1}{M}\summ \sumk \pnkm \big\{(\brktm - \brtm)-(\brkcm - \brcm)\big\}^2  = \varsigma^2_{H r} + o_p(1).
\end{eqnarray*}

From the above results, we conclude that 
$\hat{\sigma}_{ss}^2 \stackrel{P}{\to} \varsigma^2_r(\pi) + \varsigma^2_{Hr}$, hence complete the proof.
\end{proof}

\end{document}